\documentclass[twocolumn,trackchanges]{aastex631}
\pdfoutput=1 
\usepackage[T1]{fontenc}
\usepackage[figure,figure*]{hypcap}
\usepackage{amsmath,amstext,bm}
\usepackage{graphicx,textcomp,fancyhdr,hyperref}

\graphicspath{{./}{figures/}}

\DeclareRobustCommand{\okina}{%
\raisebox{\dimexpr\fontcharht\font`A-\height}{%
    \scalebox{0.8}{`}%
  }%
}
\newcommand{\ms}{m~s$^{-1}$}

\newcommand{\kms}{km~s$^{-1}$}
\newcommand{\istar}{i_\star}
\newcommand{\iorb}{i_\text{orb}}

\newcommand{\vsini}[0]{v\sin\istar}
\newcommand{\Lsun}{\text{L}_\odot}
\newcommand{\Msun}{\text{M}_\odot}
\newcommand{\Rsun}{\text{R}_\odot}
\newcommand{\Mearth}{\text{M}_\oplus}
\newcommand{\Rearth}{\text{R}_\oplus}

\newcommand{\teff}{T_\text{eff}}
\newcommand{\feh}{\ensuremath{[\mbox{Fe}/\mbox{H}]}}
\newcommand{\rphk}{\ensuremath{R'_{\mbox{\scriptsize HK}}}}
\newcommand{\lrphk}{\ensuremath{\log{\rphk}}}
\newcommand{\logg}{\ensuremath{\log{g}}}

\newcommand{\Mb}{11.1}
\newcommand{\Mberr}{1.2}
\newcommand{\Mc}{2.8}
\newcommand{\Mcerr}{2.3}
\newcommand{\Mculim}{6.4}
\newcommand{\Rb}{1.8}
\newcommand{\Rberr}{0.1}
\newcommand{\Rc}{1.6}
\newcommand{\Rcerr}{0.1}
\newcommand{\rhob}{9.9}
\newcommand{\rhobuperr}{2.1}
\newcommand{\rhobloerr}{1.7}
\newcommand{\rhoc}{3.6}
\newcommand{\rhocuperr}{3.3}
\newcommand{\rhocloerr}{3.0}
\newcommand{\rhoculim}{4.1}

\newcommand{\tess}{{\it TESS}}

\shorttitle{TKS XII: A Dense USP with an Atmosphere?}
\shortauthors{Rubenzahl, Dai et al.}

\begin{document}

\title{The \tess{}-Keck Survey. XII. A Dense $1.8~\Rearth$ Ultra-Short-Period Planet Possibly Clinging to a High-Mean-Molecular-Weight Atmosphere After the First Gyr}

\author[0000-0003-3856-3143]{Ryan A. Rubenzahl}
\altaffiliation{NSF Graduate Research Fellow}
\affiliation{Department of Astronomy, California Institute of Technology, Pasadena, CA 91125, USA}

\author[0000-0002-8958-0683]{Fei Dai} 
\altaffiliation{NASA Sagan Fellow}
\affiliation{Department of Astronomy, California Institute of Technology, Pasadena, CA 91125, USA}
\affiliation{Division of Geological and Planetary Science, California Institute of Technology, Pasadena, CA 91125, USA}

\author[0000-0001-8638-0320]{Andrew W. Howard}
\affiliation{Department of Astronomy, California Institute of Technology, Pasadena, CA 91125, USA}

\author[0000-0001-6513-1659]{Jack J. Lissauer}
\affiliation{Space Science \& Astrobiology Division, MS 245-3, NASA Ames Research Center, Moffett Field, CA 94035, USA}

\author[0000-0002-4290-6826]{Judah Van Zandt}
\affiliation{Department of Physics \& Astronomy, University of California Los Angeles, Los Angeles, CA 90095, USA}

\author[0000-0001-7708-2364]{Corey Beard}
\altaffiliation{NASA FINESST Fellow}
\affiliation{Department of Physics \& Astronomy, University of California Irvine, Irvine, CA 92697, USA}

\author[0000-0002-8965-3969]{Steven Giacalone}
\altaffiliation{NSF Astronomy and Astrophysics Postdoctoral Fellow}
\affiliation{Department of Astronomy, California Institute of Technology, Pasadena, CA 91125, USA}

\author[0000-0001-8898-8284]{Joseph M. Akana Murphy}
\altaffiliation{NSF Graduate Research Fellow}
\affiliation{Department of Astronomy and Astrophysics, University of California, Santa Cruz, CA 95064, USA}

\author[0000-0003-1125-2564]{Ashley Chontos}
\altaffiliation{Henry Norris Russell Fellow}
\affiliation{Department of Astrophysical Sciences, Princeton University, 4 Ivy Lane, Princeton, NJ 08540, USA}
\affiliation{Institute for Astronomy, University of Hawai\okina i, Honolulu, HI 96822, USA}

\author[0000-0001-8342-7736]{Jack Lubin}
\affiliation{Department of Physics \& Astronomy, University of California Irvine, Irvine, CA 92697, USA}

\author[0000-0002-4480-310X]{Casey L. Brinkman}
\affiliation{Institute for Astronomy, University of Hawai\okina i, Honolulu, HI 96822, USA}

\author[0000-0003-0298-4667]{Dakotah Tyler}
\affiliation{Department of Physics \& Astronomy, University of California Los Angeles, Los Angeles, CA 90095, USA}

\author[0000-0003-2562-9043]{Mason G.\ MacDougall}
\affiliation{Department of Physics \& Astronomy, University of California Los Angeles, Los Angeles, CA 90095, USA}

\author[0000-0002-7670-670X]{Malena Rice}
\affiliation{Department of Astronomy, Yale University, New Haven, CT 06511, USA}

\author[0000-0002-4297-5506]{Paul A.\ Dalba}
\affiliation{Department of Astronomy and Astrophysics, University of California, Santa Cruz, CA 95064, USA}

\author[0000-0002-7216-2135]{Andrew W. Mayo}
\affiliation{Department of Astronomy, University of California Berkeley, Berkeley, CA 94720, USA}

\author[0000-0002-3725-3058]{Lauren M. Weiss}
\affiliation{Department of Physics and Astronomy, University of Notre Dame, Notre Dame, IN 46556, USA}

\author[0000-0001-7047-8681]{Alex S. Polanski}
\affiliation{Department of Physics and Astronomy, University of Kansas, Lawrence, KS 66045, USA}

\author[0000-0002-3199-2888]{Sarah Blunt}
\affiliation{Department of Astronomy, California Institute of Technology, Pasadena, CA 91125, USA}
\affiliation{CIERA, Northwestern University, Evanston IL 60201}

\author[0000-0001-7961-3907]{Samuel W.\ Yee}
\affiliation{Department of Astrophysical Sciences, Princeton University, 4 Ivy Lane, Princeton, NJ 08540, USA}
\affiliation{Center for Astrophysics \textbar \ Harvard \& Smithsonian, 60 Garden St, Cambridge, MA 02138, USA}

\author[0000-0002-0139-4756]{Michelle L. Hill}
\affiliation{Department of Earth and Planetary Sciences, University of California, Riverside, CA 92521, USA}

\author[0000-0002-9751-2664]{Isabel Angelo}
\affiliation{Department of Physics \& Astronomy, University of California Los Angeles, Los Angeles, CA 90095, USA}

\author[0000-0002-1845-2617]{Emma V. Turtelboom}
\affiliation{Department of Astronomy, University of California Berkeley, Berkeley, CA 94720, USA}

\author[0000-0002-5034-9476]{Rae Holcomb}
\affiliation{Department of Physics \& Astronomy, University of California Irvine, Irvine, CA 92697, USA}

\author[0000-0003-0012-9093]{Aida Behmard}
\affiliation{Department of Astrophysics, American Museum of Natural History, 200 Central Park West, Manhattan, NY 10024, USA}

\author[0000-0001-9771-7953]{Daria Pidhorodetska} 
\affiliation{Department of Earth and Planetary Sciences, University of California, Riverside, CA 92521, USA}

\author[0000-0002-7030-9519]{Natalie M. Batalha}
\affiliation{Department of Astronomy and Astrophysics, University of California, Santa Cruz, CA 95064, USA}

\author{Ian J. M. Crossfield}
\affiliation{Department of Physics and Astronomy, University of Kansas, Lawrence, KS 66045, USA}

\author[0000-0001-8189-0233]{Courtney Dressing}
\affiliation{Department of Astronomy, University of California Berkeley, Berkeley, CA 94720, USA}

\author[0000-0003-3504-5316]{Benjamin Fulton}
\affiliation{NASA Exoplanet Science Institute/Caltech-IPAC, MC 314-6, 1200 E. California Blvd., Pasadena, CA 91125, USA}

\author[0000-0001-8832-4488]{Daniel Huber}
\affiliation{Institute for Astronomy, University of Hawai\okina i, Honolulu, HI 96822, USA}
\affiliation{Sydney Institute for Astronomy (SIfA), School of Physics, University of Sydney, NSW 2006, Australia}

\author[0000-0002-0531-1073]{Howard Isaacson}
\affiliation{Department of Astronomy, University of California Berkeley, Berkeley, CA 94720, USA}
\affiliation{Centre for Astrophysics, University of Southern Queensland, Toowoomba, QLD, Australia}

\author[0000-0002-7084-0529]{Stephen R. Kane}
\affiliation{Department of Earth and Planetary Sciences, University of California, Riverside, CA 92521, USA}

\author[0000-0003-0967-2893]{Erik A. Petigura}
\affiliation{Department of Physics \& Astronomy, University of California Los Angeles, Los Angeles, CA 90095, USA}

\author[0000-0003-0149-9678]{Paul Robertson}
\affiliation{Department of Physics \& Astronomy, University of California Irvine, Irvine, CA 92697, USA}

\author[0000-0003-3623-7280]{Nicholas Scarsdale}
\affiliation{Department of Astronomy and Astrophysics, University of California, Santa Cruz, CA 95064, USA}

\author[0000-0003-4603-556X]{Teo Mo\v{c}nik}
\affiliation{Gemini Observatory/NSF's NOIRLab, 670 N. A\okina ohoku Place, Hilo, HI 96720, USA}

\author[0000-0002-3551-279X]{Tara Fetherolf}
\affiliation{Department of Physics, California State University, San Marcos, CA 92096, USA}
\affiliation{Department of Earth and Planetary Sciences, University of California, Riverside, CA 92521, USA}

\author[0000-0002-6492-2085]{Luca Malavolta} 
\affiliation{Dipartimento di Fisica e Astronomia ``Galileo Galilei'', Universit\`a di Padova, Vicolo dell'Osservatorio 3, IT-35122, Padova, Italy}
\affiliation{INAF - Osservatorio Astronomico di Padova, Vicolo dell'Osservatorio 5, IT-35122, Padova, Italy}

\author[0000-0001-7254-4363]{Annelies Mortier} 
\affiliation{School of Physics and Astronomy, University of Birmingham, Edgbaston, Birmingham B15 2TT, UK}

\author[0000-0002-4272-4272]{Aldo Fiorenzano} 
\affiliation{Fundaci\'on Galileo Galilei - INAF, Rambla Jos\'e Ana Fernandez P\'erez 7, E-38712 Bre\~na Baja, Tenerife, Spain}

\author[0000-0002-5752-6260]{Marco Pedani} 
\affiliation{Fundaci\'on Galileo Galilei - INAF, Rambla Jos\'e Ana Fernandez P\'erez 7, E-38712 Bre\~na Baja, Tenerife, Spain}

\begin{abstract}

The extreme environments of ultra-short-period planets (USPs) make excellent laboratories to study how exoplanets obtain, lose, retain, and/or regain gaseous atmospheres. We present the confirmation and characterization of the USP TOI-1347~b, a $\Rb \pm \Rberr~\Rearth$ planet on a 0.85~day orbit that was detected with photometry from the \tess{} mission. We measured radial velocities of the TOI-1347 system using Keck/HIRES and HARPS-N and found the USP to be unusually massive at $\Mb \pm \Mberr~\Mearth$. The measured mass and radius of TOI-1347~b imply an Earth-like bulk composition. A thin H/He envelope (>0.01\% by mass) can be ruled out at high confidence. The system is between 1 and 1.8 Gyr old; therefore, intensive photoevaporation should have concluded. We detected a tentative phase curve variation (3$\sigma$) and a secondary eclipse (2$\sigma$) in \tess{} photometry, which if confirmed could indicate the presence of a high-mean-molecular-weight atmosphere. We recommend additional optical and infrared observations to confirm the presence of an atmosphere and investigate its composition.
\end{abstract}



\section{Introduction}

Ultra-short-period planets, or USPs, are exoplanets that orbit their stars with short orbital period ($< 1$~day). USPs tend to not exceed 2~$\Rearth$ in size, save for the closest of the hot Jupiter (HJ) population. This ``hot-Neptune desert'' \citep{Mazeh2016} has been hypothesized to be sculpted by mass loss mechanisms, such as photoevaporation \citep{OwenWu2017} or core-powered mass loss \citep{Gupta2019}. Such mechanisms destroy the atmospheres of smaller planets, whereas giant planets can better resist mass loss \citep[in fact, HJ atmospheres can become inflated;][]{Batygin2011, Grunblatt2017}. The \textit{Kepler} mission \citep{Borucki2010} took the first steps to map out the demographics of close-in transiting planets and revealed that USPs exist around $\sim$0.5--0.8\% of GK stars \citep{SanchisOjeda2014}. Another key discovery from \textit{Kepler} was the ``radius gap'' around 1.7--1.9~$\Rearth$, which separated the bimodal peaks corresponding to the smaller super-Earth population (no atmosphere) and the larger sub-Neptunes ($\gtrsim$1\% H/He atmosphere) \citep{FGap2017, FGap2018}. 

To date, only two non-giant USPs have been discovered with radii larger than 2~$\Rearth$. TOI-849~b \citep{Armstrong849} is a massive ($\sim$40~$\Mearth$) rocky world that is likely the stripped core of a giant planet (perhaps via giant collisions). LTT-9779~b \citep[29~$\Mearth$, 0.79 days;][]{Jenkins} defies its environment with a substantial (9\% by mass) H/He atmosphere and a 4.6~$\Rearth$ radius. This suggests some USPs can retain atmospheres. If so, where is the boundary between bare rocky cores and those with residual (or secondary) atmospheres? 

A close-in orbiting planet will reflect starlight and emit thermal radiation, causing the observed brightness of the system to vary with the planet's orbital position. This variation, called the phase curve, depends on the planet's albedo and day-night temperature contrast. Phase curve variations have been used to directly probe the surfaces of several USPs, finding some to be bare rock \citep[LHS 3844~b;][]{LHS3844,Kane2020} and others to perhaps possess high mean-molecular weight atmospheres \citep[55 Cnc~e;][]{Demory55cnce_phasecurve}.

Unlike \textit{Kepler}, which observed a single patch of sky for four years, the Transiting Exoplanet Survey Satellite \citep[\tess{};][]{TESS} is an all-sky transit survey that has discovered thousands of close-in planets around nearby bright stars that are amenable to radial velocity (RV) follow-up measurements. RVs provide key insight into the existence of exoplanetary atmospheres by measuring the planet's mass, thereby constraining the bulk density, planet surface gravity (which is related to atmospheric scale height), and its ability to resist photoevaporation. Our collaboration, the \tess{}-Keck Survey \citep[TKS;][]{TKS0}, has been monitoring 86 \tess{} systems with RV follow-up using the HIgh Resolution Echelle Spectrometer \citep[HIRES;][]{HIRES} at the Keck-I Telescope. As part of TKS, we have measured the masses of several USPs, such as TOI-561~b \citep{Weiss561} and TOI-1444~b \citep{Dai_1444}, both of which have rocky compositions. In this paper, we present the discovery and mass measurement of the USP TOI-1347~b. The heaviest of the non-giant USPs\footnote{Other than the exceptional LTT-9779~b and TOI-849~b, which likely belong to a separate class of planets than the $\lesssim 10~\Mearth$ USPs \citep{Dai_1444}} to date, TOI-1347~b shows hints of phase curve variability and a secondary eclipse in its \tess{} photometry, which may indicate the presence of a high mean-molecular-weight atmosphere. We also detected a second transiting planet in the system at 4.84~d, TOI-1347~c, in line with the common trend for USPs to have nearby outer companions which can shepherd the migrations of their USPs into their present sub-day orbits \citep{Millholland_USP}.

This paper is structured as follows. In Section~\ref{sec:stparams} we characterize the host star using spectroscopy. We rule out stellar companions using speckle imaging in Section~\ref{sec:speckle}. In Section~\ref{sec:tess} we analyze the \tess{} light curve to measure the stellar rotation period, planetary transits, and tentative phase-curve variability and secondary eclipse. In Section~\ref{sec:rvs} we present our RV measurements and the resulting mass constraints for both planets. Finally, we discuss the implications of our observations for the TOI-1347 system in Section~\ref{sec:discussion}.

\section{Host Star Properties}\label{sec:stparams}

\subsection{Spectroscopic Properties}\label{sec:specprops}
TOI-1347 (TIC 229747848) is a late G type ($\teff = 5464 \pm 100$~K) star that is relatively active ($\lrphk = -4.66$), showing strong variability in both photometry (see Section~\ref{sec:tess}) and RVs (see Section~\ref{sec:rvs}). 

We obtained high-resolution spectra of TOI-1347 with HIRES (with the B3 decker, $R = 67,000$) and HARPS-N ($R = 115,000$); see Section~\ref{sec:rvs} for details. We applied the \texttt{SpecMatch-Synthetic}~\citep{Petigura2015} algorithm to the HIRES spectrum to measure $\teff$, $M_\star$, Fe/H, and $\log g$. These parameters, as well as the J and K magnitudes from the TICv8 catalog \citep{ticv82} and the \textit{Gaia} parallax \citep{gaiadr3}, were input into \textit{isoclassify}~\citep{Huber17, Berger2020} to constrain $R_\star$. We also used \texttt{KeckSpec} (Polanski in prep) to compute alpha elemental abundances. Stellar parameters for TOI-1347 are included in the catalog of \citet{MacDougall2023}. We verified that our values agree with the values therein to within 1$\sigma$.

We similarly derived spectroscopic parameters from the co-added HARPS-N spectra of TOI-1347 using the FASMA spectral synthesis package \citep{Tsantaki2018, Tsantaki2020}. We verified that all derived quantities were in agreement with the HIRES values. In particular, the HARPS-N data are of sufficient resolution to detect $\vsini$, whereas with HIRES we only obtain an upper limit. Table~\ref{tab:stellar} lists our adopted stellar parameters.

\subsection{Age}
\begin{figure}
    \centering
    \includegraphics[width=0.495\textwidth]{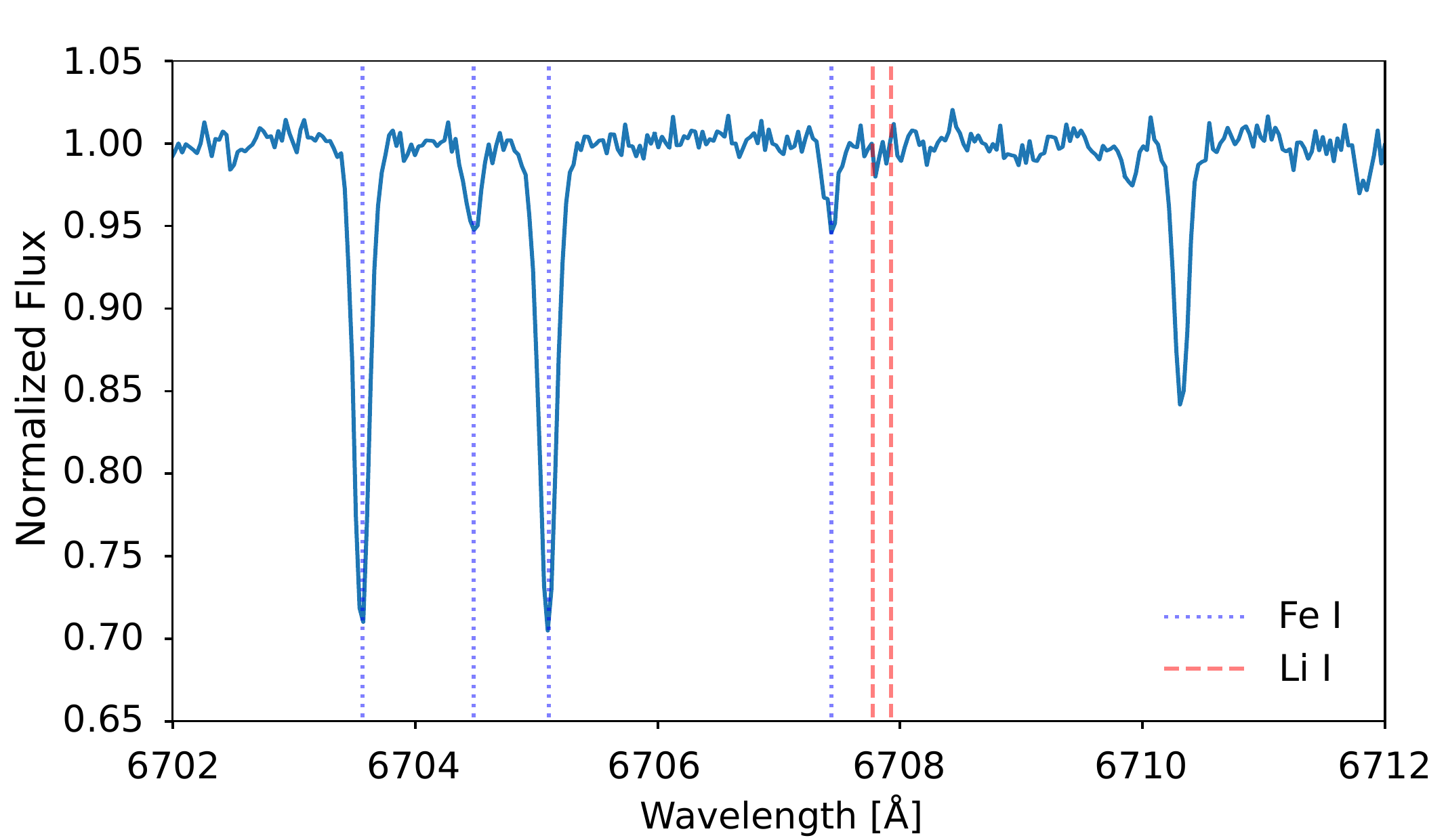}
    \caption{HIRES spectrum of TOI-1347 in the neighborhood of the lithium doublet. Nearby Fe~I lines are labelled. No absorption attributed to lithium was detected.}
    \label{fig:lithium}
\end{figure}

We constrained the age of the host star using several methods. Using a stellar rotation period of $16.1 \pm 0.3$ days (see Section~\ref{sec:rotation} for details), we estimated the gyrochronological age of the star. The scaling relation of \citet{Mamajek} gives an age of $1.33\pm0.06$~Gyr. Using the latest empirical relations of \citet{Bouma_2023}, the age is $1.7\pm0.1$~Gyr. We also obtained an age estimate using chromospheric activity in the Ca~II~H\,\&\,K lines from our HIRES spectra. We measured $\log R_{\rm HK}^\prime = -4.66 \pm 0.05$ using the method of \citet{Isaacson}. Combining with the calibration of \citet{Mamajek}, the corresponding age is $1.6\pm0.4$~Gyr. We also looked for the lithium doublet in our HIRES spectra, but were unable to detect the lithium doublet above the noise floor in the continuum (see Fig. \ref{fig:lithium}). We placed an upper limit of 2~m$\textup{\AA}$ (95\% confidence) on the equivalent width. According to \citet{Berger_li}, the star is consistent with field stars and is most likely older than the Hyades cluster ($\sim$650~Myr). Lastly, \citet{MacDougall2023} derived an age of $0.8_{-0.6}^{+1.1}$~Gyr from isochronal fitting. 

The age uncertainties above do not account for systematic errors. Therefore, we combined age indicators with an unweighted mean. We adopted a wide age uncertainty of 0.4 Gyr to reflect the systematic uncertainties between the different age estimators. Our best age estimate for TOI-1347 is thus $1.4 \pm 0.4$~Gyr. Importantly, this age is longer than the timescale on which photoevaporation operates, which is typically confined to the first few hundred Myr when the star is active in X-rays and extreme UV \citep{Ribas, Tu2015}.

\begin{figure}
    \centering
    \includegraphics[width=0.495\textwidth]{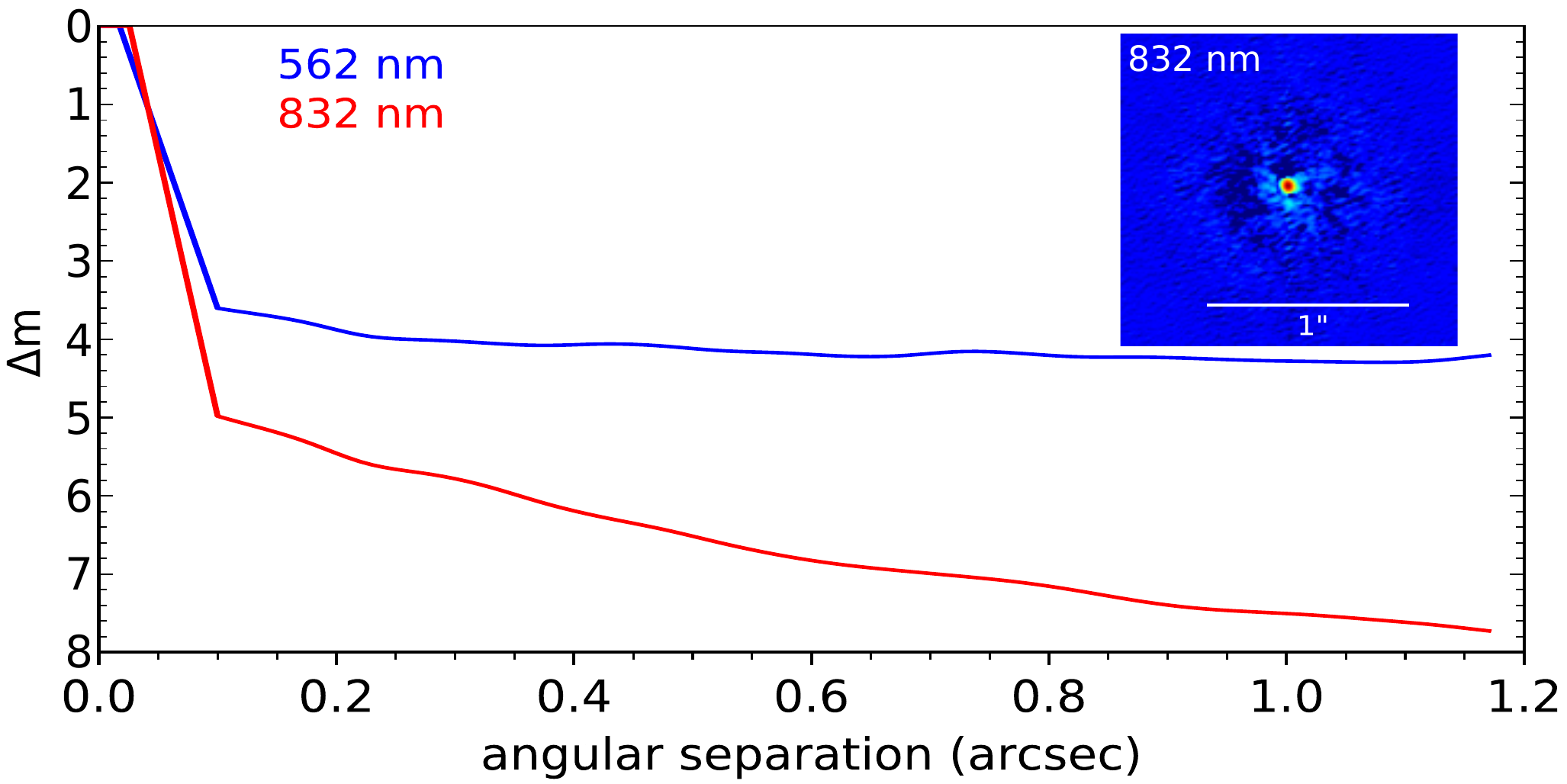}
    \caption{Contrast curves around TOI-1347 from Gemini/{\okina}Alopeke; the inset shows the reconstructed image at 832~nm.  No companions are detected. 
    \label{fig:speckle}}
\end{figure}

\section{High Resolution Imaging}\label{sec:speckle}
To help validate the transiting planets, we observed TOI-1347 with the {\okina}Alopeke \citep{Alopeke} dual-channel speckle imaging instrument on Gemini-N (PI: Crossfield) with a pixel scale of 0.01 arcsec/pixel and a full width at half maximum resolution of 0.02 arcsec. With {\okina}Alopeke we obtained simultaneous speckle imaging at 562~and 832~nm, with a total of seven observing blocks each consisting of one thousand 60~ms exposures.

We processed these images with the speckle pipeline of \citet{howell:2011}, which yielded the 5-sigma sensitivity curves and reconstructed image shown in Fig.~\ref{fig:speckle}. The curves do not show companions at angular separations of 0.5~arcsec or greater at a contrast of 4.12 mag at 562 nm and 6.52 mag at 832 nm (Fig.~\ref{fig:speckle}).

\begin{figure*}
    \centering
    \includegraphics[width=\textwidth]{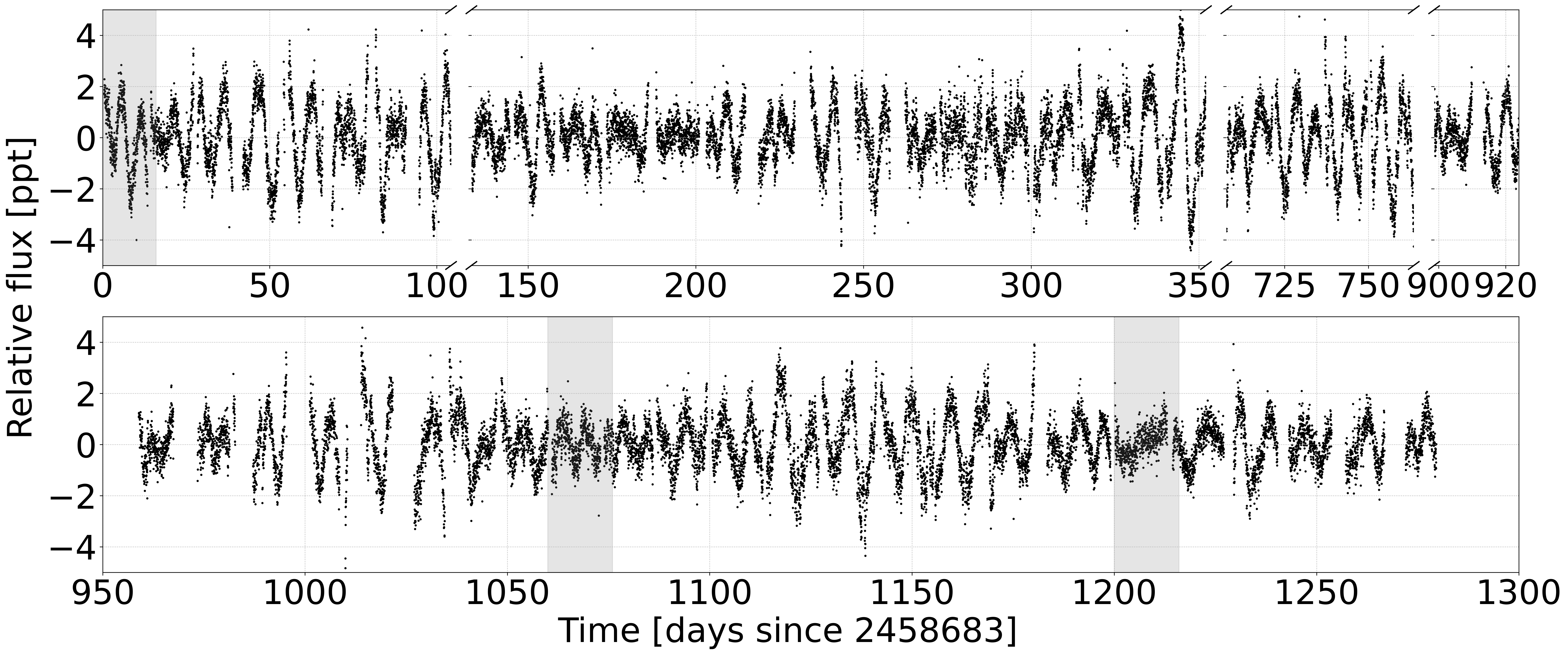}
    \includegraphics[width=\textwidth]{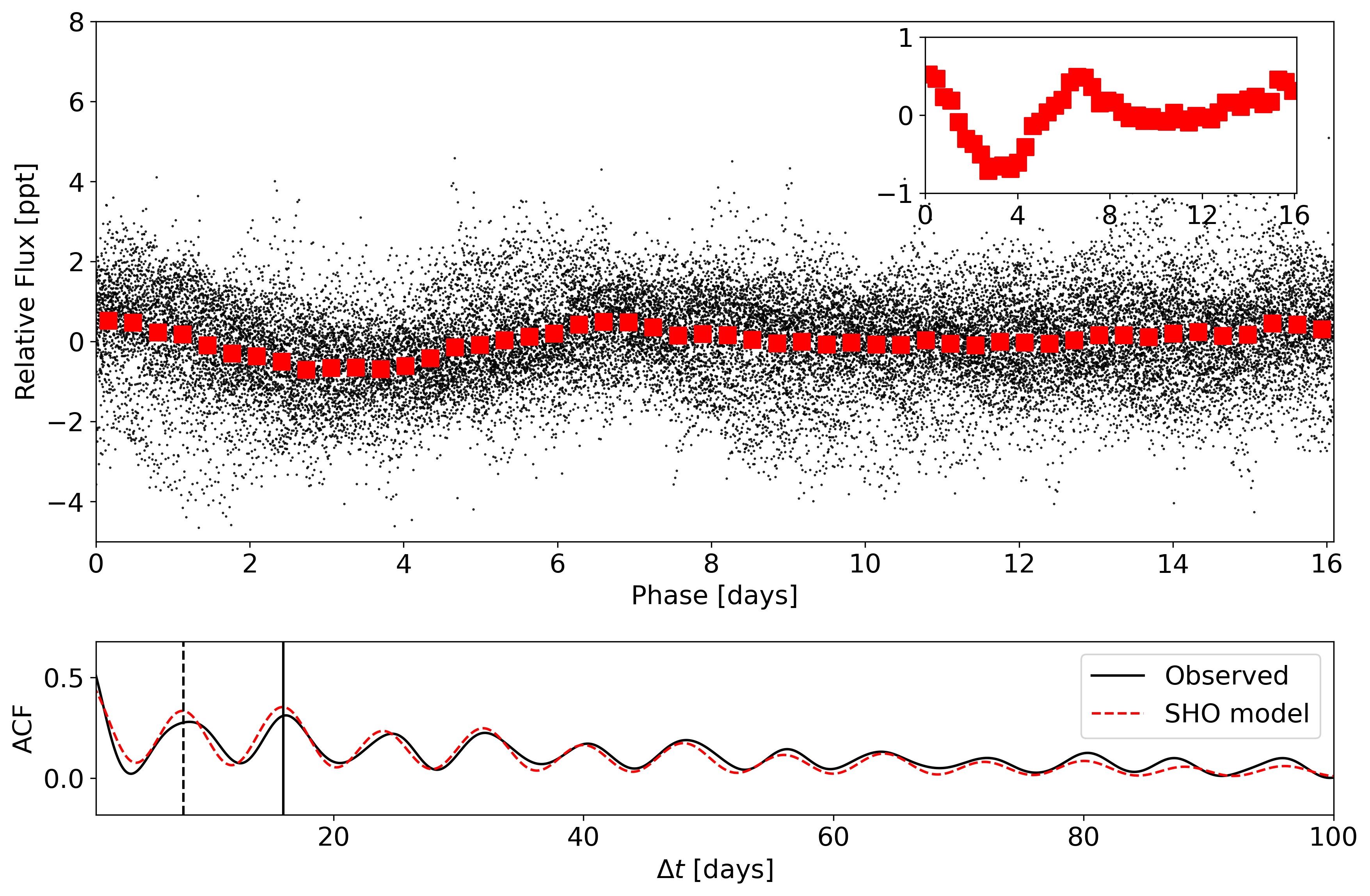}
    \caption{\textbf{Top:} The full 120~s cadence SPOC \tess{} light curve, binned to 30~min. Rotationally modulated variability is strong and evolves over time; the three shaded regions highlight example 16~day windows in which different numbers of maxima/minima are observed. \textbf{Middle:} The 30-min light curve phased to the 16.1~day rotation period. The red points further bin the folded data to $\sim 8$~hour bins. The inset zooms in on these binned phased data and highlights the tendency for every other set of maximum/minimum to repeat in amplitude. \textbf{Bottom:} The ACF of the photometry, showing regular peaks at all multiples of 8~days (dashed line) with the highest peak at 16 days (dark line). The red dashed line shows the best-fit SHO model described in Section~\ref{sec:rotation}.}
    \label{fig:lightcurve}
\end{figure*}

\startlongtable 
\begin{deluxetable*}{lcccc}
\tablecaption{Stellar Parameters of TOI-1347 \label{tab:stellar}}
\tablehead{\colhead{Parameter} & \colhead{Value} & \colhead{Unit} & \colhead{Source}}
\startdata
TIC ID & TIC 229747848 & & TIC v8.2$^a$  \\[-0.05cm]
Right Ascension & 18:41:18.4 & hh:mm:ss &  TIC v8.2$^a$ \\[-0.05cm]
Declination & +70:17:24.19 & dd:mm:ss & TIC v8.2$^a$ \\[-0.05cm]
V magnitude & $11.168 \pm 0.013$ & & TIC v8.2$^a$ \\[-0.05cm]
TESS magnitude & $10.7157 \pm 0.0066$ & & TIC v8.2$^a$ \\[-0.05cm]
J magnitude & $10.011\pm0.02$ & & TIC v8.2$^a$ \\[-0.05cm]
K magnitude & $9.616 \pm 0.015$ & & TIC v8.2$^a$ \\[-0.05cm]
\textit{Gaia} magnitude & $11.2076 \pm 0.0005$ & & \textit{Gaia} DR3$^b$ \\[-0.05cm]
Parallax & $6.7803 \pm 0.0112$ & mas & \textit{Gaia} DR3$^b$ \\[-0.05cm]
RA proper motion & $-6.0883 \pm 0.0150$ & mas/yr & \textit{Gaia} DR3$^b$ \\[-0.05cm]
Dec proper motion & $26.1020 \pm 0.0150$ & mas/yr & \textit{Gaia} DR3$^b$ \\
Galactic U & $-5.31 \pm 0.09$ & km/s & Derived from \textit{Gaia} DR3 astrometry$^c$ \\
Galactic V & $-7.72 \pm 0.46$ & km/s & Derived from \textit{Gaia} DR3 astrometry$^c$ \\
Galactic W & $4.91 \pm 0.23 $ & km/s & Derived from \textit{Gaia} DR3 astrometry$^c$ \\
\hline
Luminosity & $0.55 \pm 0.02$ & $\Lsun$ &  Isoclassify$^d$ \\[-0.05cm]
Radius & $0.83 \pm 0.03$ & $\Rsun$ &  Isoclassify$^d$ \\[-0.05cm]
Mass & $0.913 \pm 0.033$ & $\Msun$ &  SpecMatch-Synthetic$^e$ \\[-0.05cm]
$\teff$ & $5464 \pm 100$ & K & SpecMatch-Synthetic$^e$ \\[-0.05cm]
$\logg$ & $4.64 \pm 0.10$ & &  SpecMatch-Synthetic$^e$  \\[-0.05cm]
$\feh$ & $0.04 \pm 0.06$ & dex & SpecMatch-Synthetic$^e$ \\[-0.05cm]
$[\alpha/\text{Fe}] $ & $-0.03 \pm 0.06$ & dex & KeckSpec$^f$ \\[-0.05cm]
$\vsini$ & $< 3$ & km/s & SpecMatch-Synthetic$^e$  \\[-0.05cm]
$\vsini$ & $2.9 \pm 0.1 $ & km/s & FASMA$^g$  \\[-0.05cm]
$v_{eq}$  & $2.61\pm0.11$ & km/s & From $P_\text{rot}$ and $R_\ast$ \\[-0.05cm] 
$\lrphk$ & $-4.66 \pm 0.05$ &  & From Keck/HIRES spectrum \\ 
\hline
$P_\text{rot}$ & $16.1 \pm 0.3$ & days & From ACF of TESS photometry (Section~\ref{sec:rotation}) \\[-0.05cm]
$P_\text{rot}$ & $16.3 \pm 0.6$ & days & From \texttt{TESS-SIP} with TESS photometry (Section~\ref{sec:rotation}) \\[-0.05cm]
$P_\text{rot}$ & $16.2 \pm 0.3$ & days & From GP fit to RVs (Section~\ref{sec:rvmodel}) \\
\hline 
Age & $1.7 \pm 0.1$ & Gyr & From $P_\text{rot}$ using empirical relations of \citet{Bouma_2023} \\
Age  & $1.33 \pm 0.06$ & Gyr & From $P_\text{rot}$ using empirical relations of \citet{Mamajek} \\
Age  & $1.6 \pm 0.4$ & Gyr & From $\lrphk$ using empirical relations of \citet{Mamajek} \\
Age  & $0.8_{-0.6}^{+1.1}$ & Gyr & From isochronal fitting by \citet{MacDougall2023} \\
Age  & $> 650$ & Myr & From Li and comparison to Hyades \citep{Berger_li} \\
Age  & $1.4 \pm 0.4 $& Gyr & Adopted value \\
\hline
\enddata
\tablenotetext{}{
$^a$\citealt{ticv82}. 
$^b$\citealt{gaiadr3}. 
$^c$Using local standard of rest from \citealt{LSR}, as implemented in \texttt{PyAstronomy.pyasl.gal\_uvw} \citep{PyAstronomy}.  
$^d$\citealt{Huber17}, using the SpecMatch-Synthetic results from a HIRES spectrum. Uncertainties are random errors for the adopted model grid. To account for systematic errors, see \citet{Tayar2022}. 
$^e$\citealt{Petigura2015}, using a HIRES spectrum taken at $R\sim67000$ with no iodine. 
$^f$Polanski (in prep), using a HIRES spectrum taken at $R\sim67000$ with no iodine. 
$^g$\citealt{Tsantaki2018, Tsantaki2020}, using the co-added HARPS-N spectra. 
}
\end{deluxetable*}

\section{\tess{} Photometry}\label{sec:tess}

TOI-1347 was observed by \tess{} in sectors 14--26, 40, 41, and 47--60, which span UT Jul 18 2019 to Jan 18 2023. The Science Processing Operations Center \citep[SPOC;][]{jenkinsSPOC2016} detected two transiting planet candidates which were subsequently diagnosed and vetted as \tess{} Objects of Interest (TOIs) 1347.01 (b) and 1347.02 (c) \citep{Guerrero}.

We downloaded the 20~s and 120~s cadence SPOC light curves using \texttt{lightkurve} \citep{lightkurve}. We removed all data points with a non-zero Quality Flag, i.e. those suffering from cosmic rays or other known systematic issues. We then stitched and normalized the multi-sector data using \texttt{lightkurve} and applied a 5-$\sigma$ sigma-clipping. \texttt{lightkurve} also provides the correction for scattered light (2\% contamination reported by ExoFOP). The resulting light curve, shown in Figure~\ref{fig:lightcurve}, exhibits significant but coherent periodic variability corresponding to rotationally modulated surface inhomogeneities on the stellar disk.

\subsection{Stellar Rotation Period}\label{sec:rotation}

\begin{figure}
    \centering
    \includegraphics[width=0.495\textwidth]{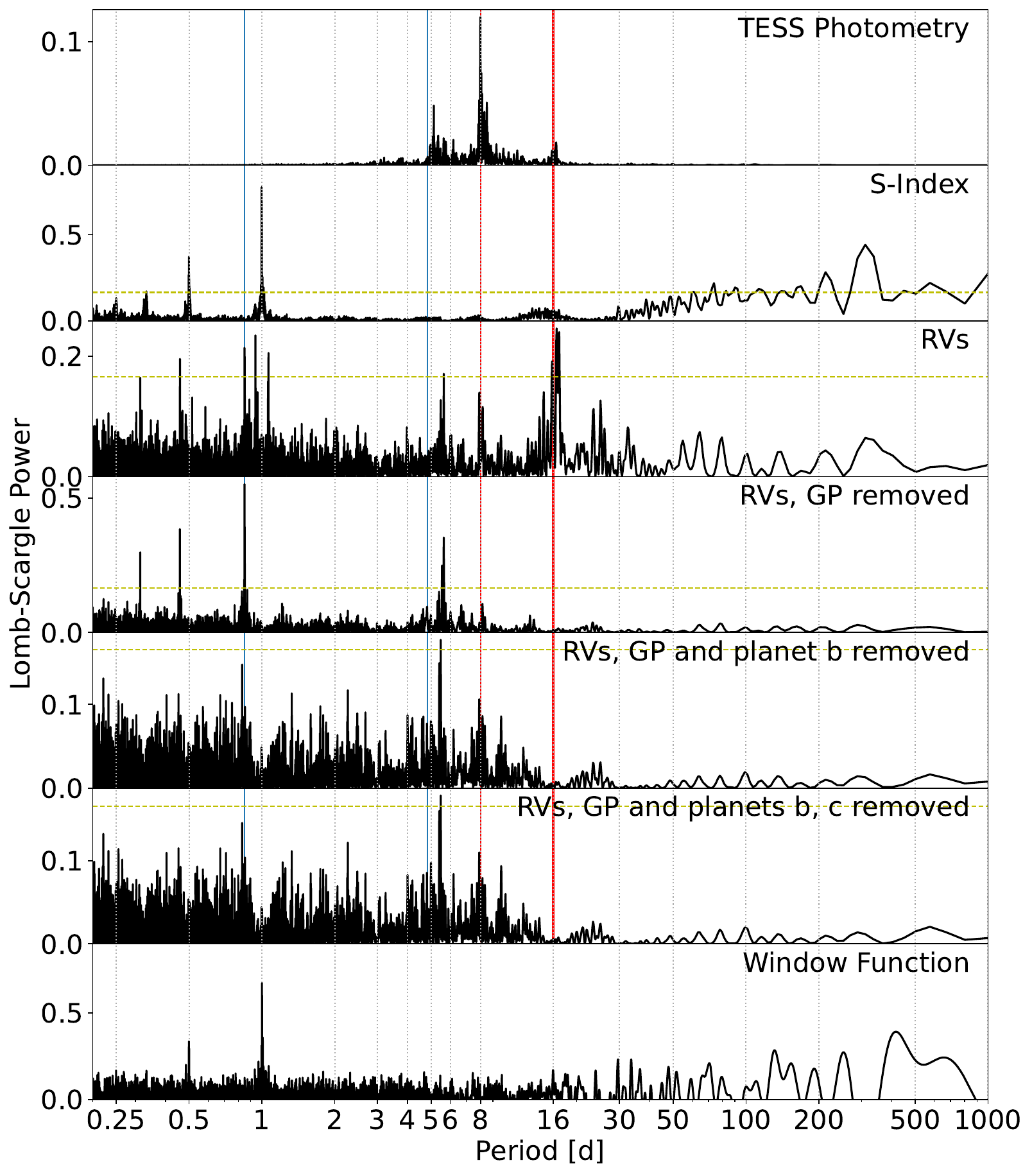}
    \caption{Lomb-Scargle periodograms of, from top to bottom, the \tess{} photometry, S-Indices, RVs, RVs with the GP model (Section~\ref{sec:rvmodel}) removed, the GP-corrected RVs with the Keplerian model for the USP subtracted, the GP-corrected RVs with both planets subtracted, and the window function \citep{DawsonFabrycky2010} of the RV time series. The periodograms are computed using \texttt{astropy.timeseries.LombScargle} \citep{astropy}. The blue dashed lines correspond to the orbital periods of TOI-1347 b and c, and the two dark red lines are drawn at 8~days (thin) and 16~days (thick). The horizontal yellow line is the 1\% false alarm probability.}
    \label{fig:periodogram}
\end{figure}

The effect of a starspot on photometry is to reduce the observed (integrated) intensity when on the visible hemisphere. The net effect of many spots is quasiperiodic variability that can be treated as time-correlated noise \citep{Haywood2014, Rajpaul2015, Aigrain2023}. The \tess{} photometry of TOI-1347 shows strong rotational variability with maxima/minima occurring roughly every $\sim 8$~days (see the top panel in Figure~\ref{fig:lightcurve}). A Lomb-Scargle periodogram of the photometry shows a strong peak at 8~days (Fig~\ref{fig:periodogram}). However, a closer examination of the light curve reveals that the depths of adjacent maxima/minima are dissimilar. In fact, the depths of alternating maxima/minima tend to have similar amplitudes. This can be explained if the star has a 16~day rotation period and multiple spot groups concentrated on opposite hemispheres of the star, an effect noted for a number of other stars \citep{spinspotter}. In fact, a periodogram analysis of the RV dataset shows a strong peak at 16~days (Fig~\ref{fig:periodogram}). Spots affect RVs in much the same way as photometry by breaking the flux balance across star's rotational velocity profile~\citep{Saar1997}, in addition to suppressing the convective blueshift \citep{Haywood2016}.

While the periodogram is essentially a Fourier decomposition showing the amplitude of the best-fitting sine wave at all possible periods, the autocorrelation function (ACF) measures the self-similarity of a time series as a function of time delay. Thus, if the variability in a time series is not sinusoidal (asymmetric, more complex shape), then the ACF will give a better estimate of the periodicity than the periodogram. The ACF of the \tess{} photometry has its highest peak at $\sim$16~days (Fig~\ref{fig:lightcurve}) with adjacent ACF peaks alternating between high and low amplitude. An analysis of the ACF using \texttt{spinspotter} \citep{spinspotter} successfully identified the half-period effect by checking that the odd peaks are less than 10\% the height of the even peaks. \texttt{spinspotter} returns a rotation period of $P_\text{rot} = 16.1 \pm 0.3$~days by averaging the locations of the even-numbered peaks only. We also measured the stellar rotation period using the TESS Systematics Insensitive Periodogram algorithm \citep[TESS-SIP][]{TESS_SIP}. \texttt{TESS-SIP} simultaneously corrects for instrument systematics while performing the periodogram search on the TESS Simple Aperture Photometry. Using \texttt{TESS-SIP} for sectors 14-26 for TOI-1347, we measured a stellar rotation period of $16.34 \pm 0.57$~days. 

With all of this considered, we adopt the $\sim$16~day solution as the rotation period of TOI-1347 and explain the Lomb-Scargle periodogram peak at $P_\text{rot}/2$ as arising from antipodal spot groups. The persistence of repeated peaks every 8~days in the ACF, even out to beyond 100~days, can be explained if the starspots on TOI-1347 live for many rotation periods. Following the prescription of \citet{Giles2017}, we fit the observed ACF with an underdamped simple harmonic oscillator (uSHO), including a second component with power at $P/2$:
\begin{equation}\label{eq:usho}
    y = e^{-\Delta t/\tau} \left[A \cos\left(\frac{2\pi \Delta t}{P}\right) + B\cos\left(\frac{2\pi \Delta t}{P/2}\right) + y_0, \right].
\end{equation}
In Eq.~\ref{eq:usho}, $y$ is the ACF strength and the coefficients $A$ and $B$ give the relative strengths of the antipodal spot groups. We used \texttt{scipy.optimize} \citep{scipy} to find the maximum a-posteriori (MAP) solution, then used that as a seed for a Markov-chain Monte Carlo (MCMC) analysis using \texttt{emcee}~\citep{emcee}. The fit recovered $P = 16.0$~d, and the best-fit exponential decay timescale was $\tau = 45$~days, roughly three times the rotation period. Both coefficients $A = 0.05$ and $B = 0.18$ were constrained to nonzero values, again supporting the multiple spot group hypothesis.

Lastly, our \texttt{SpecMatch-Synthetic} analysis of the HIRES spectrum in Section~\ref{sec:specprops} yielded only an upper limit corresponding to the line spread function of HIRES (roughly 2.2~{\kms}). \citet{Masuda2022} recently showed that on the population level, such nondetections of $\vsini$ are most consistent with a $< 3$~{\kms} upper-limit. Combining $P_\text{rot} = 16.1$~d with the stellar radius from Section~\ref{sec:specprops}, we get an equatorial rotational velocity of $v_{eq} = 2.6 \pm 0.1$~{\kms}. This is consistent with a $<3$~{\kms} upper-bound for HIRES, though it is slightly smaller than the measured HARPS-N value. Were the rotation period 8~d, we would instead have $v_{eq} \sim 5$~{\kms}, which (barring a misaligned stellar inclination of $\lesssim 30^\circ$) would be detectable in the HIRES spectrum and inconsistent with the HARPS-N measurement. If astrophysical, the slightly larger $\vsini$ could be explained by differential rotation and long-lived starspots at higher latitudes.

\subsection{Additional Transiting Planets?}\label{sec:transit_search}

We searched the \tess{} light curve for planetary transits with a Box-Least-Squares algorithm \citep[BLS,][]{Kovac2002} as described in \citet{Dai_1444}. We recovered the two planet candidates reported by ExoFOP at 0.84 and 4.84 days. We did not find any other transit signals with SNR~$> 6.5$.

\subsection{Transit Modeling}\label{sec:transit}
Our transit analysis closely follows that described in \citep{Dai_1444}. Briefly, we generated transit signals using the {\tt Python} package {\tt Batman} \citep{Kreidberg2015}, parameterized with the stellar density ($\rho_\ast$) to break the degeneracy between the scaled semimajor axes ($a/R_\ast$) and impact parameters ($b$) of the transiting planets \citep{Seager}. We imposed a prior using the best-fit stellar density from our adopted $M_\ast$ and $R_\ast$ of $\rho_\ast = 2.25 \pm 0.26$~g~cm$^{-3}$. For the limb darkening coefficients, we used the prior and parameterization scheme of \citet{Kipping} for a quadratic limb darkening law ($q_1$ and $q_2$). The other transit parameters included the orbital period ($P_{\text{orb}}$), time of conjunction ($T_{\text{c}}$), planet-to-star radius ratios ($R_{\text{p}}/R_\star$), scaled orbital distances ($a/R_\star$, computed from stellar density and orbital period), orbital inclinations (cos$i$), orbital eccentricities ($e$), and the arguments of pericenter ($\omega$). We initially allowed non-zero eccentricities for both TOI-1347~b and c; however, the posterior distributions are fully consistent with circular orbits. Neither the existing transit nor RV data (Section~\ref{sec:rvs}) support the detection of nonzero eccentricities. In our final fits, we chose to restrict both planets to circular orbits to reduce model complexity.

%

Our transit fitting pipeline takes the following steps:
\begin{enumerate}
    \item Fit a global model of all transit epochs, assuming no transit timing variations (TTVs). This model is found by maximizing the likelihood with the {\tt Levenberg-Marquardt} method implemented in {\tt Python} package {\tt lmfit} \citep{LM}. 
    \item Search for TTVs: We held the transit shape parameters fixed and fit for the mid-transit times of each transit epoch. We removed out-of-transit variations with a quadratic model. We did not detect significant TTVs for either planet, so we continued with fixed orbital periods. 
    \item Full model. We sampled the posterior distributions of both planets jointly using the MCMC framework implemented in \textit{emcee} \citep{emcee} with 128 walkers initialized near the MAP solution. We ran \texttt{emcee} for 50000 steps, ensuring that this was many times longer than the autocorrelation of various parameters (typically 100s of steps). 
\end{enumerate}

\begin{figure}
    \centering\includegraphics[width=0.47\textwidth]{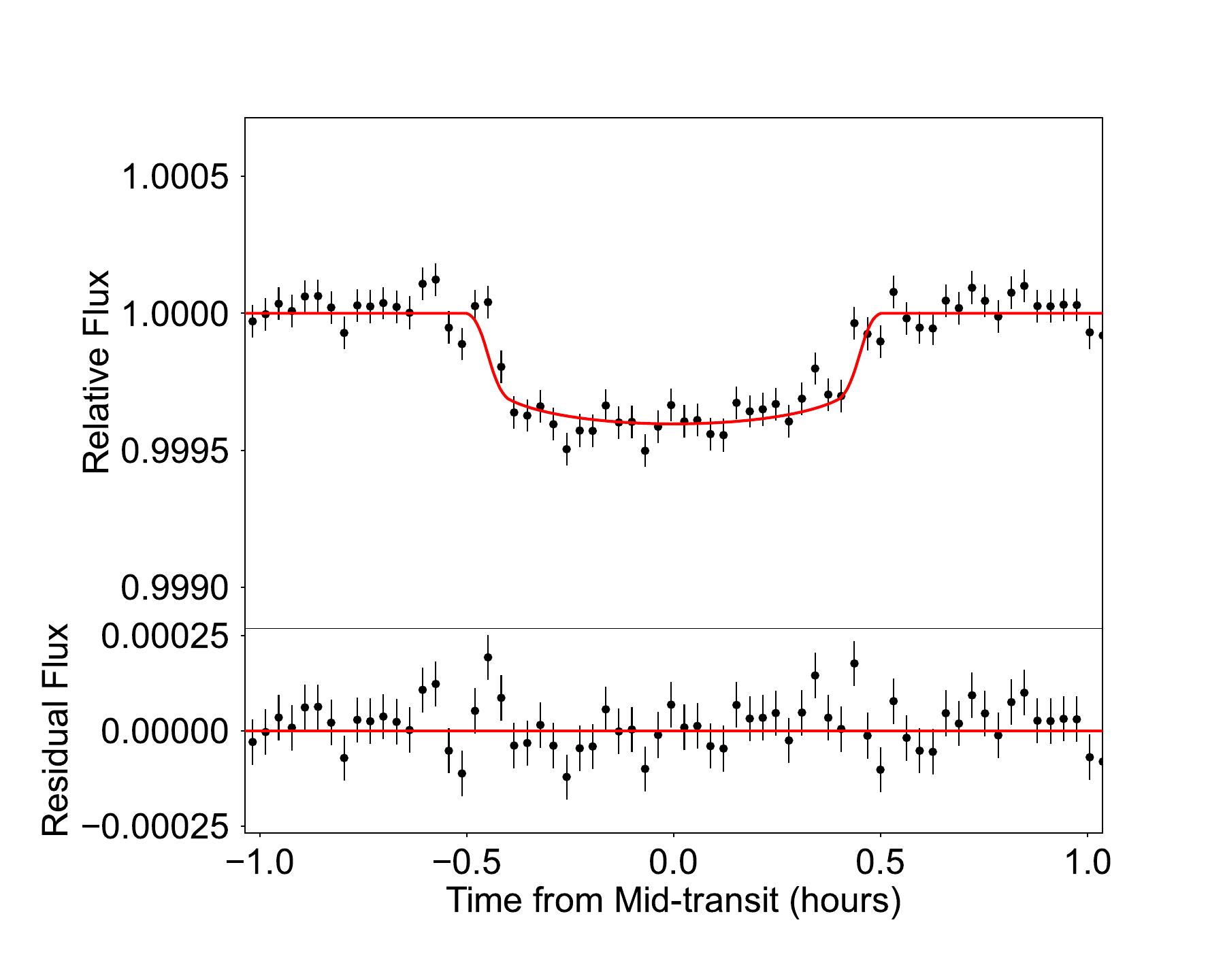}
    \includegraphics[width=0.47\textwidth]{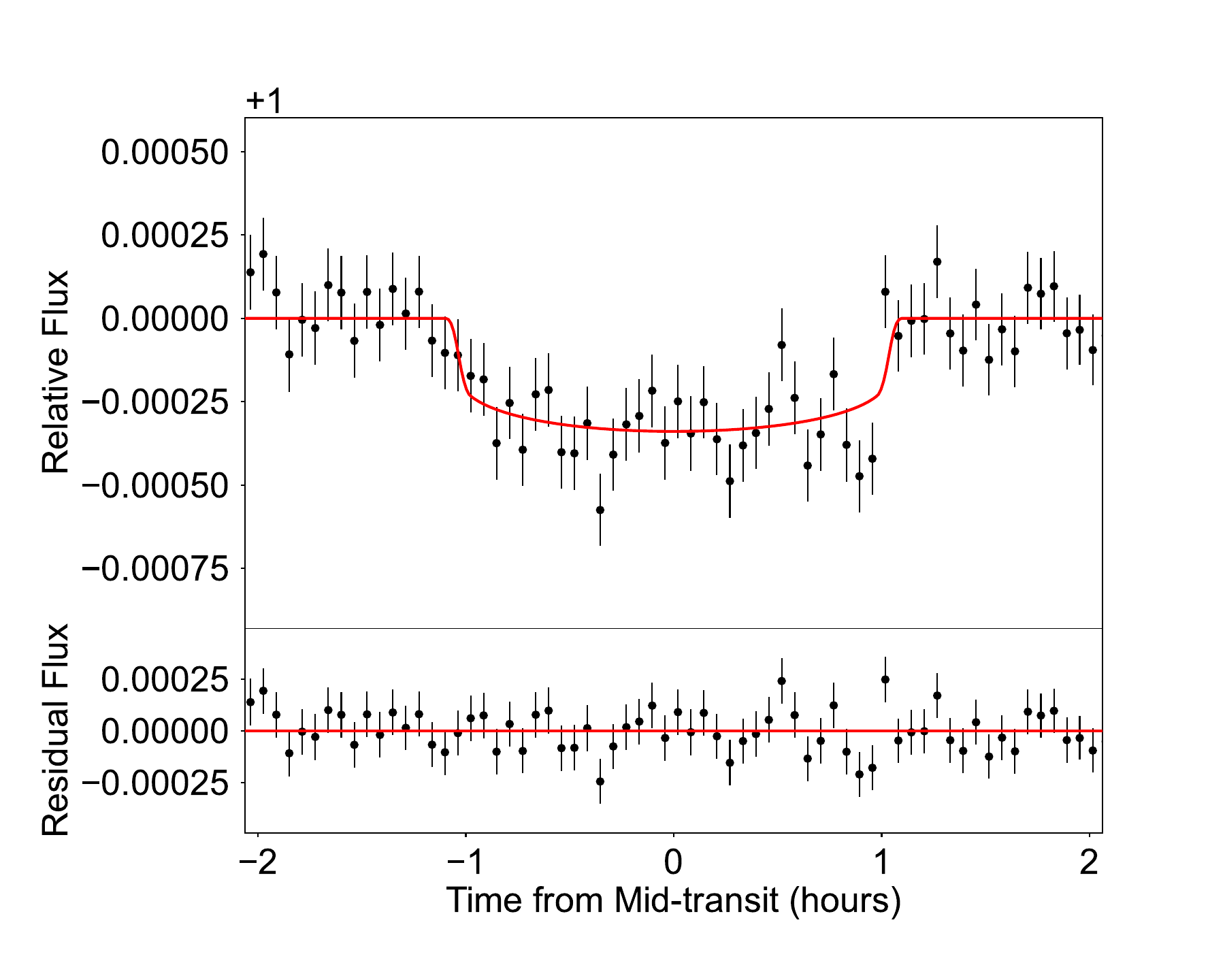}
    \caption{The 2~min \tess{} light curves phase-folded and binned using the orbital periods of TOI-1347~b (top) and c (bottom). The best-fit transit models are shown with red solid lines.}
    \label{fig:transit_fit}
\end{figure}

Fig. \ref{fig:transit_fit} shows the phase-folded and binned transits of TOI-1347~b and c with the MAP model. Using the stellar radius derived in Section~\ref{sec:stparams}, we derived $R_{p,b} = \Rb \pm \Rberr~\Rearth$ and $R_{p,c} = \Rc \pm \Rcerr~\Rearth$, in agreement with the radii measured by \citet{MacDougall2023} ($R_{p,b} = 1.81^{+0.09}_{-0.06}~\Rearth$ and $R_{p,c} = 1.68^{+0.09}_{-0.07}$).

\startlongtable
\begin{deluxetable*}{lllll}
\tablecaption{Transit and RV Parameters of the TOI-1347 System} 
\label{tab:planet_para}
\tablehead{
\colhead{Parameter}  & \colhead{Symbol} & \colhead{Prior} &  \colhead{Posterior Distribution} & \colhead{Unit}}
\startdata
\textbf{TOI-1347~b} \\
Planet/Star Radius Ratio & $R_p/R_\star$  & $\text{Uniform}(-10, 10)$ on $\ln(R_p/R_\star)$ & $0.02039\pm0.00072$ &   \\
Time of Conjunction & $T_c$  & $\text{Uniform}(1682, 1683)$ & $1682.71214\pm0.00060$ & (BJD-2457000)  \\
Orbital Period & $P_\text{orb}$  & $\text{Uniform(-10,10)}$ on $\ln(P_\text{orb})$ & $0.84742346\pm0.00000061$ & days \\
Orbital Inclination & $\iorb$  & $\text{Uniform}(-1, 1)$ on $\cos\iorb$ & $79.5\pm0.7$ & deg  \\
Orbital Eccentricity  & $e$  & - & 0 (fixed) &  \\
Impact Parameter & $b$  & Derived & $0.82\pm0.03$ & \\
Scaled Semi-major Axis & $a/R_\star$  & Derived & $4.43\pm 0.21$ &  \\
RV Semi-amplitude  & $K$ & $\text{Normal}(0, 50)$  &$7.74^{+0.80}_{-0.79}$ & {\ms} \\
Planetary Radius   & $R_{\rm p}$ & Derived  & $ \Rb \pm \Rberr$ & $\Rearth$ \\
Planetary Mass  & $M_{\rm p}$ & Derived  & $\Mb \pm \Mberr$ & $\Mearth$ \\
Bulk Density  & $\rho$ & Derived  & $\rhob_{-\rhobloerr}^{+\rhobuperr}$ & g cm$^{-3}$ \\
Equilibrium Temperature & $T_{\rm eq}$ & Derived ($A_B = 0.7$)  & $1400 \pm 40$ & K \\
\hline
\textbf{TOI-1347 c} \\
Planet/Star Radius Ratio & $R_p/R_\star$  & $\text{Uniform}(-10, 10)$ on $\ln(R_p/R_\star)$  & $0.0179\pm0.0010$ &   \\
Time of Conjunction & $T_c$  & $\text{Uniform}(1678,1679)$ & $1678.5059\pm0.0021$ & (BJD-2457000)  \\
Orbital Period  & $P_\text{orb}$  & $\text{Uniform}(-10,10)$ on $\ln(P_\text{orb})$ & $4.841962\pm0.000012$ & days \\
Orbital Inclination & $i$  & $\text{Uniform}(-1,1)$ on $\cos\iorb$ & $87.5\pm0.4$ & deg  \\
Orbital Eccentricity  & $e$  & - & 0 (fixed) &  \\
Impact Parameter & $b$  & Derived & $0.73\pm0.08$ & \\
Scaled Semi-major Axis & $a/R_\star$  & Derived & $14.18\pm 0.49$ &   \\
RV Semi-amplitude  & $K$ & $\text{Normal}(0, 50)$  &$1.08^{+0.91}_{-0.92}~(< 2.59~\text{at 95\%})$ & {\ms} \\
Planetary Radius   & $R_{\rm p}$ & Derived  &$\Rc \pm \Rcerr$ & $\Rearth$ \\
Planetary Mass  & $M_{\rm p}$ & Derived  & $\Mc \pm \Mcerr ~(< \Mculim~\text{at 95\%}) $ & $\Mearth$ \\
Bulk Density  & $\rho$ & Derived  & $\rhoc_{-\rhocloerr}^{+\rhocuperr} ~(< \rhoculim~\text{at 95\%}) $ & g cm$^{-3}$ \\
Equilibrium Temperature & $T_{\rm eq}$ & Derived ($A_B = 0.1$)  & $1000 \pm 25$ & K \\
\hline
\textbf{Stellar Parameters} \\
Stellar Density & $\rho_\star$ & $\text{Normal}(2.25, 0.26)$ &$ 2.30\pm0.24 $ & g~cm$^{-3}$\\
Limb Darkening  & $q_1$ & $\text{Uniform}(0,1)$  &$0.28 \pm 0.20$ & \\
Limb Darkening  & $q_2$ & $\text{Uniform}(0,1)$  &$0.35 \pm 0.28$ & \\
HIRES RV Jitter     & $\sigma_{\rm HIRES}$  & $\text{HalfNormal}(0, 10)$  &$4.05^{+0.73}_{-0.67}$ & {\ms} \\
HARPS-N RV Jitter   & $\sigma_{\rm HARPS}$  & $\text{HalfNormal}(0, 10)$  &$8.20^{+2.67}_{-1.98}$ & {\ms} \\
GP Amplitude        & $A_\text{GP}$    & $\text{InverseGamma(1, 5)}$ & $9.70^{+1.16}_{-1.06}$ & {\ms} \\
Rotation Period     & $P_\text{GP}$         & $\text{Normal}(\ln(16), 0.5)$ on $\ln(P_\text{GP})$ & $16.16^{+0.35}_{-0.35}$ & days \\
Oscillator Quality Factor    & $Q_0$      & $\text{HalfNormal}(0, 2)$ on $\ln(Q_0)$ & $1.49^{+0.74}_{-0.91}$ & \\
Quality Factor Difference   & $\Delta Q$ & $\text{Normal}(0, 2)$ on $\ln(\Delta Q)$ & $1.71^{+1.32}_{-2.33}$ & \\
Fractional Amplitude & $f$  & $\text{Uniform}(0.1, 1)$ & $0.73^{+0.19}_{-0.24}$ & \\   
\hline
\textbf{Instrumental RV Parameters} \\
RV Offset (HIRES)   & $\gamma_{\rm HIRES}$ & $\text{Normal}(0, 200)$  &$-2.12^{+0.90}_{-0.87}$   & {\ms} \\
RV Offset (HARPS-N) & $\gamma_{\rm HARPS}$ & $\text{Normal}(0, 200)$ &$6.29^{+3.35}_{-3.41}$    & {\ms} \\
\enddata
\end{deluxetable*}

\subsection{Phase Curve}\label{sec:phase curve}

We searched the \tess{} light curve for phase curve variations and secondary eclipses of TOI-1347 b. First, we masked the in-transit data and removed long-term variability (stellar and/or instrumental) using the method of \citet{Sanchis2013}. This involves fitting a linear function of time to the out-of-transit data points within a window of 2$\times$ the orbital period, then dividing the best-fit function within that window, repeating for every data point in the light curve. This detrended light curve is then phase-folded to the orbital period of TOI-1347~b.

\begin{figure}
    \centering
    \includegraphics[width=0.495\textwidth]{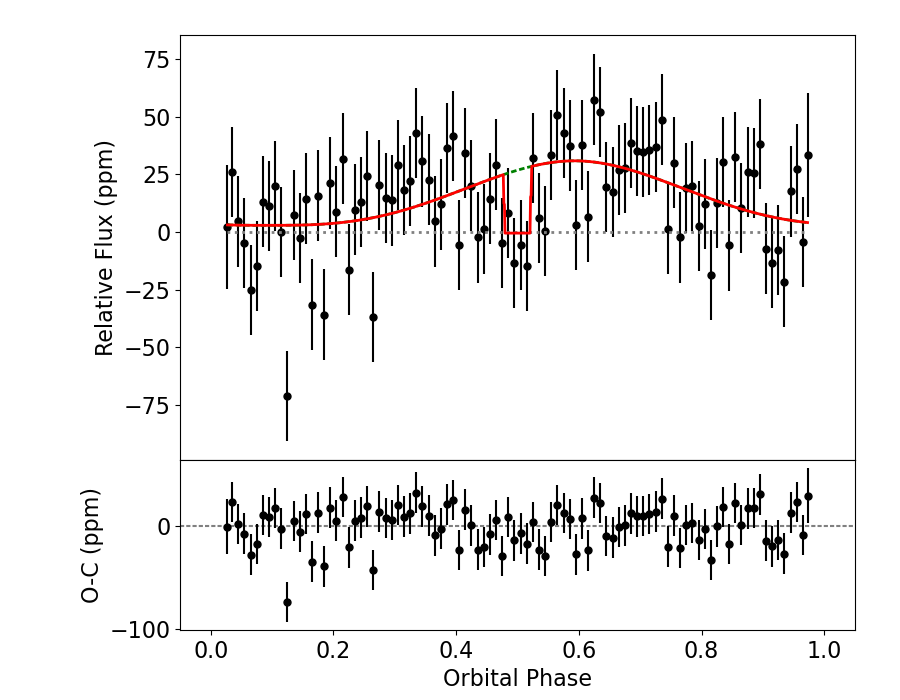}
    \includegraphics[width=0.49\textwidth]{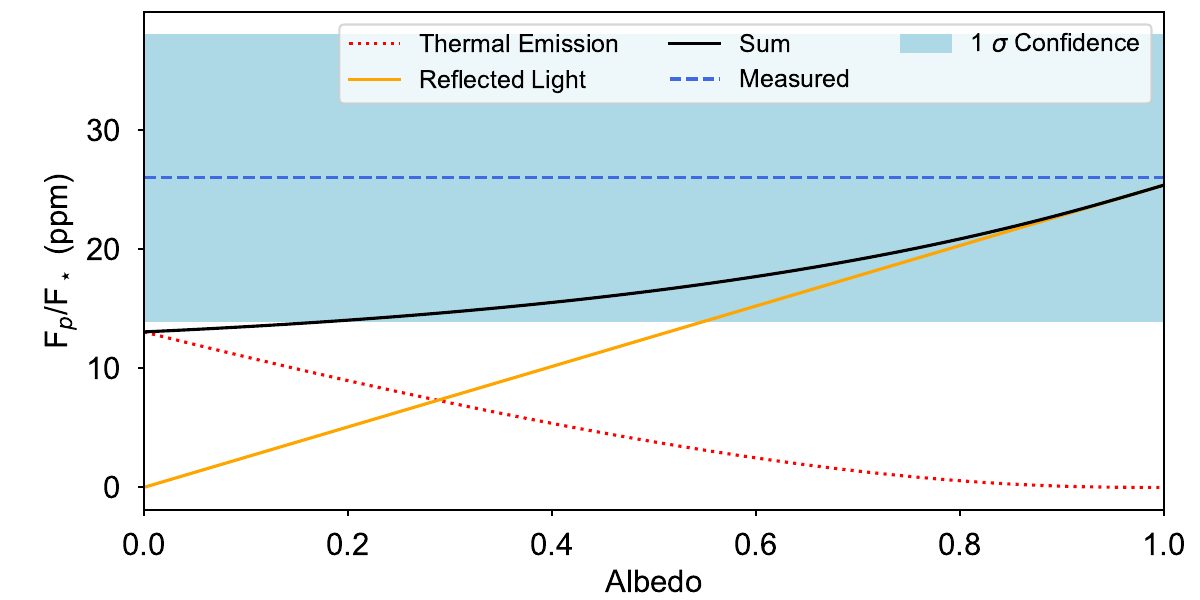}
    \caption{{\bf Upper}: The phase curve and secondary eclipse of TOI-1347~b as observed by \tess{}. The best-fit model is shown by the red curve. The green dotted line gives the model with no eclipse. The phase curve is detected at the 2$\sigma$ level. It is likely a combination of thermal emission and reflected light in the \tess{} band (600--1000~nm). {\bf Lower}: The thermal emission (red dotted line) and reflected light (orange solid line) from TOI-1347~b as a function of the Bond albedo. The blue dashed line and shaded area are the measured secondary eclipse depth ($F_p/F_\star$) and its 1$\sigma$ central interval.}
    \label{fig:phase_curve}
\end{figure}

The resultant phase curve is shown in Figure~\ref{fig:phase_curve}. We were able to detect a tentative phase curve variation (3$\sigma$) and a secondary eclipse (2$\sigma$) using a joint model. To model the secondary eclipse, we simply modified the best-fit transit model by shifting the mid-transit times by half the orbital period (i.e., assuming $e=0$) and turning off limb darkening ($q_1$ = $q_2$ = 0). The secondary eclipse depth ($\delta_{\rm sec}$) is allowed to vary freely to account for a combined effect of reflected stellar light and thermal emission from the planet's night side. For the phase curve, we used a Lambertian disk model \citep[see e.g.][]{Demory55cnce_phasecurve} parameterized by an amplitude $A$ and phase offset of the peak $\theta$. We sampled the posterior distribution with a MCMC analysis similar to that described in Section \ref{sec:transit}. We found a secondary eclipse depth of $\delta_{\rm sec}= 26\pm12$~ppm and a phase curve amplitude of $A = 28\pm 9$~ppm. The peak of the phase curve variation is shifted by $33\pm14^{\circ}$ to the west of the planet.

The lower panel of Figure \ref{fig:phase_curve} compares the amount of thermal emission vs.\ reflected light as a function of the planet's Bond albedo ($A_B$). At high albedo, the planet is more reflective and the equilibrium temperature will be lower. In this case, the phase curve will be dominated by reflected stellar light rather than thermal emission. This is necessary to explain the large secondary eclipse depth measured in the \tess{} band. Moreover, we marginally detected a phase offset of  $33\pm14^{\circ}$ to the west of the planet. Both of these effects can be explained if TOI-1347~b is retaining a high-mean-molecular-weight atmosphere. Silicate clouds in this atmosphere could produce a high albedo, while partial cloud coverage may produce the observed phase offset to the West \citep[see e.g. Kepler-7 b, ][]{Demory2013}. However, the data in hand are insufficient to definitively confirm the presence of an atmosphere on TOI-1347~b. Higher SNR follow-up observations with JWST are required.

\section{Radial Velocities}\label{sec:rvs}

\begin{figure*}
    \centering
    \includegraphics[width=0.85\textwidth]{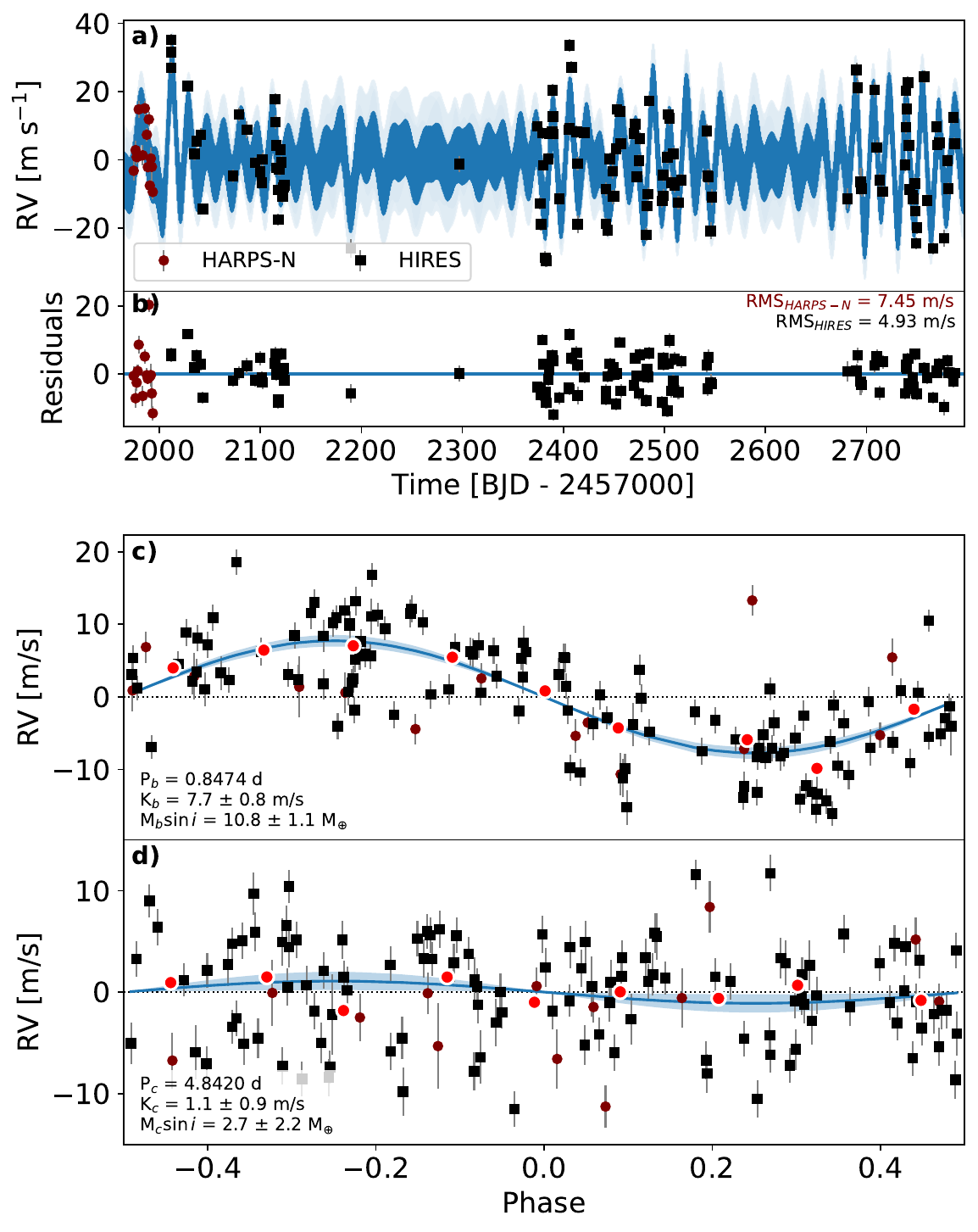}
    \caption{The adopted radial velocity model. Panel \textbf{a)} shows the HARPS-N and HIRES RV datasets, with the MAP RV model (Keplerian + GP and 1$\sigma$ uncertainty) overplotted in blue. Panel \textbf{b)} shows the residuals between the data and the MAP RV model. Panels \textbf{c)} and \textbf{d)} show the data phase-folded to the orbital period of planets b and c, respectively, with contributions from the other planet and the GP removed. The red points are equal RV bins spanning 0.1 in phase. The median and central 68\% CI of each Keplerian model is plotted in blue. The MCMC posteriors for the recovered semiamplitude and derived $M \sin i$ are also summarized in the lower-left annotations. Note that we do not include the MAP stellar jitter in the plotted errorbars; errorbars are drawn only as the measurement uncertainties to highlight the degree of unexplained scatter (i.e., jitter), given by the annotated residual RMS, to which the stellar jitter fits.}
    \label{fig:rvfit}
\end{figure*}

We collected 120 high-resolution optical spectra of TOI-1347 between UT November 28, 2019 and UT July 26, 2022 with HIRES on the Keck-I telescope as part of TKS \citep{TKS0}. We took exposures using the ``C2'' decker ($R = 45,000$) and integrated until the exposure meter reached 60,000 counts (signal-to-noise ratio (S/N) $\sim 100$ per reduced pixel) which resulted in a typical exposure time of 648~sec. We used the standard procedures of the California Planet Search \citep[CPS;][]{Howard2010} to reduce the HIRES spectra and extract precise RVs using the iodine cell for wavelength calibration~\citep{Butler1996}. The average Doppler precision per measurement was 1.83~{\ms}. 

We also obtained 14 spectra with the High-Accuracy Radial-velocity Planet Searcher in the North, installed at the Telescopio Nazionale Galileo \citep[HARPS-N;][]{HARPSN}. Observations were taken with 1800~s exposure times (average S/N is 42.3 in order 50) with simultaneous wavelength calibration provided by the Fabry-Perot etalon. Cross-correlation functions (CCFs) were created using the ESPRESSO G2 mask, and RVs were extracted by fitting for the CCF centroid \citep{Dumusque2021}. Before jointly fitting with the HIRES RVs, we subtracted the median RV from the HARPS-N dataset.

See Table~\ref{tab:rvdata} for the full RV dataset, which includes Mount Wilson S-Index values derived from the Ca~II~H\& K lines \citep{Sindex} for the HIRES data. The $S_\mathrm{HK}$ index shows a weaker power excess around $P_\text{rot}$ in a Lomb-Scargle periodogram than the RVs (Fig.~\ref{fig:periodogram}) but are primarily dominated by a 1~day sampling alias. The $S_\mathrm{HK}$ indices are correlated with the RVs at the 3.5$\sigma$ level with a Pearson correlation coefficient of 0.31 (p-value of 0.0005). This weak correlation is likely why our models in Section~\ref{sec:rvmodel} that were trained on the S-Index time series did not improve the overall fit.

\subsection{RV Model}\label{sec:rvmodel}

We chose to adopt the same two-component SHO model for our RV model that we used to model stellar variability in photometry. We used the \texttt{exoplanet} \citep{exoplanet} package to construct a Gaussian Process (GP) stellar activity model with a kernel defined by the built-in \texttt{RotationTerm} parameterization. This kernel is a mixture of two SHOs analogous to the first and second cosine terms in Eq.~\ref{eq:usho}. We also tried a GP with the quasiperiodic kernel implemented in \texttt{radvel} \citep{radvel}, both untrained and trained on the S-Indices. We found that the quasiperiodic kernel struggled to identify a single primary period, often jumping between 8~days and 16~days. In order for the MCMC to converge, a reasonably strong prior ($\pm 0.5$~d or less) had to be placed on one of these period solutions. We found this undesirable compared to the SHO kernel in \texttt{exoplanet}, which was able to lock onto the 16~day period even with wide priors of $> 10$~d. We also found the final fit to be statistically indistinguishable whether the GP was trained on the S-Indices or not, so for our final model we opted for an untrained SHO GP.

For the full mathematical definition of the SHO GP kernel, the interested reader is directed to \citet{celerite}. In brief, the kernel is parameterized by the GP amplitude ($A_\text{GP}$), the primary period of variability ($P_\text{GP}$), the quality factor of the primary mode $Q_0$, the difference in quality factors between the period and half-period modes ($\Delta Q_0$), and lastly the fractional amplitude between the two modes ($f$). We followed the guidance of the \texttt{exoplanet} tutorials to choose the appropriate priors on these parameters, which are tabulated in Table~\ref{tab:planet_para}. 

In practice, the Keplerian models are only parameterized by the RV semiamplitude $K$, which we place a wide uninformative Gaussian prior on. We do not restrict $K$ to positive values only, as such a prior can bias mass estimates to higher values, especially in the case of non-detections \citep{Weiss2014}. We imposed Gaussian priors on the periods and times of conjunction using the best-fit mean and standard deviation derived from our transit fits (Table~\ref{tab:planet_para}), effectively fixing them to the tight transit constraints. Like the transit model, we fixed the eccentricity ($e$) and argument of periastron ($\omega$) to zero. We included separate jitter terms for HIRES ($\sigma_\text{HIRES}$) and HARPS-N ($\sigma_\text{HARPS}$), and likewise separate RV offsets ($\gamma_\text{HIRES}$) and HARPS-N ($\gamma_\text{HARPS}$).

We initialized our model at the best-fit values from photometry, where applicable, and determined initial guesses for the RV semiamplitudes using \texttt{exoplanet.estimate\_semi\_amplitude}. We then solved for the MAP solution using \texttt{scipy.optimize.minimize}. The MAP parameters were then used as a seed for a MCMC exploration of the posterior. We employed a Hamiltonian Monte Carlo \cite[HMC;][]{hmc:duane87, neal12} implemented in \texttt{PyMC3} \citep{pymc3}, specifically the No-U-Turn Sampler \citep[NUTS;][]{nuts:hoffman14}. HMC and NUTS are generally more efficient than the traditional Metropolis-Hatings algorithm \citep{metropolis53, hastings70}, resulting in well-mixed MCMC chains in far fewer samples. We sampled with NUTS in four parallel chains for 5000 ``tuning'' steps, which are discarded. The number of tuning steps was chosen to obtain an acceptance fraction near the target of 0.9 which balances number of retained samples and efficient exploration of the posterior space. Each chain then collects 5000 samples for a total of 20000 posterior samples. We ensured adequate statistical independence amongst these samples by requiring that the per-parameter $\hat{R}$ statistic \citep{vehtari21} be $< 1.001$.

Our best-fitting RV model is shown in Figure~\ref{fig:rvfit}. The GP robustly recovers a primary period of $16.2 \pm 0.3$~days, even with a wide prior, independently verifying our assessment of the stellar rotation period from photometry. The USP TOI-1347~b is also robustly detected at consistent semiamplitudes, regardless of the activity model used (trained/untrained, quasi-periodic/SHO). The resulting mass of TOI-1347~b is $\Mb \pm \Mberr~\Mearth$. The second planet, TOI-1347~c, is not detected in the RVs. We adopt an upper limit of $< \Mculim~\Mearth$ at 95\% confidence. The residual RMS to the combined Keplerian + GP model is $4.9$~{\ms} for HIRES and $7.5$~{\ms} for HARPS-N, which is (expectedly) similar to the fitted stellar jitter values ($4.0 \pm 0.7$~{\ms} and $8.2_{-2.0}^{+2.7}$~{\ms} respectively), but larger than the per-measurement uncertainty of each instrument (2.5 and 1.9~{\ms}).

Of note is the large jitter for the HARPS-N RVs. We suspect this is due to the GP being primarily conditioned on the HIRES RVs, resulting in poor predictive accuracy for activity during the timespan of the HARPS-N data, which occur about 1--2 rotation cycles before the HIRES data. As a result, the jitter term for HARPS-N is inflated to compensate. We tried separate GPs for both datasets, sharing all hyperparameters except for the GP amplitude ($A_\text{GP}$). We found a nearly identical result (in fact, the best-fit HARPS-N jitter was higher) with statistically indistinguishable planet parameters, likely due to the relatively few HARPS-N data points. As a result, we adopt the single GP model, but encourage further investigation into the nature of stellar activity on TOI-1347. As it stands, there are too few HARPS-N data points to condition independent GPs, but the single GP model is potentially overfitted to the HIRES points, reducing its out-of-sample predictive accuracy \citep{Blunt2023}.

\section{Discussion}\label{sec:discussion}

\subsection{A Heavy Core Pushing the Limit of Photoevaporation}\label{sec:compositions}

\begin{figure*}
    \centering
    \includegraphics[width=\textwidth]{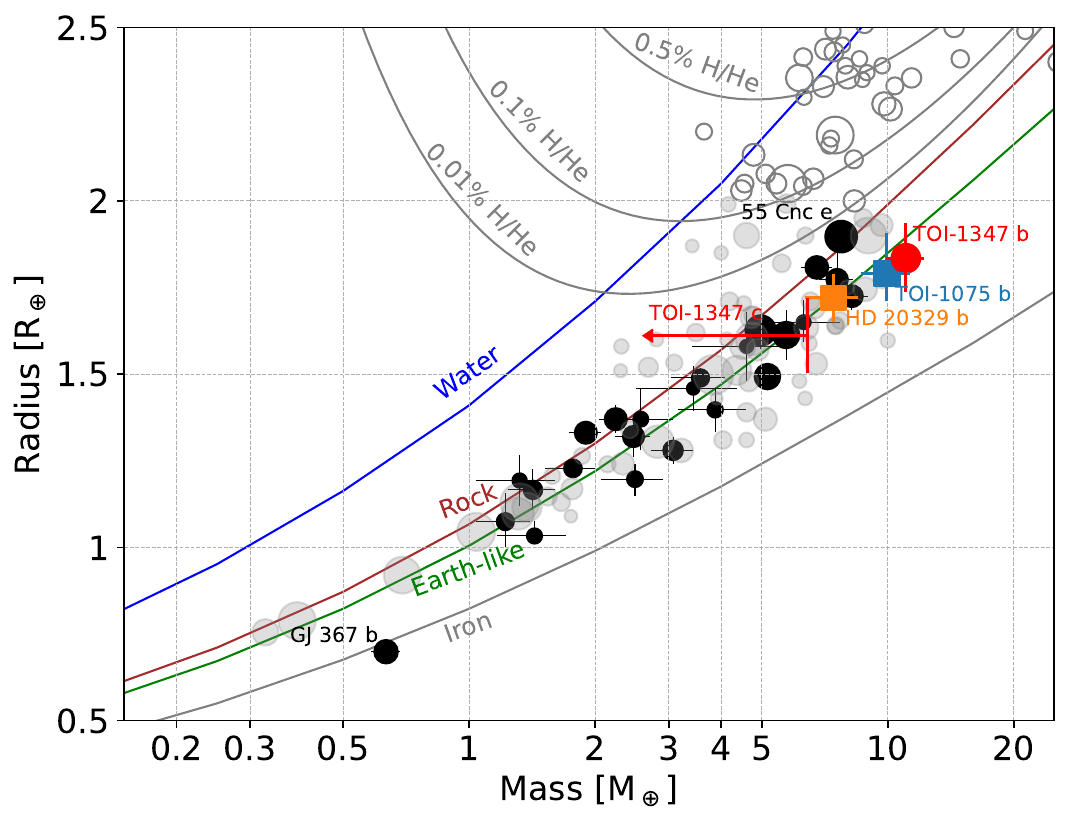}
    \caption{Mass-radius diagram of known super-Earths ($R_p < 2~\Rearth$, filled circles) and sub-Neptunes ($4~\Rearth > R_p \geq 2~\Rearth$, empty circles) with 5$\sigma$ or better mass measurements, obtained from the NASA Exoplanet Archive \citep{neadoi}. Bold fill denotes USPs. Contours from \citet{Zeng2016} are drawn for pure-iron, Earth-like (30\% iron, 70\% rock), pure-rock, and pure-water compositions. Contours from \citet{Chen_Rogers} are also drawn for 0.5\%, 0.1\%, and 0.01\% H/He envelopes surrounding rocky-composition cores, at an age of 1.4-Gyr-old and at the maximum insolation flux of 400~S$_\oplus$ for their model grids; it is worth noting that TOI 1347~b (1400~K, $A=0.7$) receives an insolation flux of around 3000~S$_\oplus$. Our mass-radius constraints for TOI-1347~b and c (95\% upper limit in mass) are plotted and labelled in red. The size of each point is proportional to $M / \sigma_M$. TOI-1347~b is the most massive super-Earth USP to date, while TOI-1347~c is smaller but likely also rocky.}
    \label{fig:mass-radius}
\end{figure*}

TOI-1347~b is the largest (in both mass and radius) of the super-Earth USPs to date. It seems to be rocky in composition and similar in iron core mass fraction to the Earth. Figure~\ref{fig:mass-radius} shows the two planets in the context of known exoplanets on the mass-radius diagram. We modeled the core composition of TOI-1347~b assuming a simple two-layer model with an iron core and a silicate (``rock'') mantle \citep{Zeng2016}. Our mass and radius measurements suggest an iron core mass fraction of 41$\pm$ 27\%, not far from Earth's 33\% core mass fraction. TOI-1347~b joins a group of well-characterized USP planets \citep{Dai2019} that are consistent with an Earth-like composition. 

With a core mass larger than 10~$\Mearth$, TOI-1347~b, along with the USPs TOI-1075~b \citep{toi1075}, and HD 20329~b \citep{hd20329}, are close to the theoretical limit for runaway accretion \citep[see, e.g.,][]{Rafikov}. How did these planets evade runaway accretion and not become gas giants? \citet{Lee_accretion} and \citet{Chachan} both noted that the local hydrodynamic conditions, the envelope opacity, and the timescale of core assembly relative to disk dissipation could all contribute to quenching runaway accretion. As more of these systems are discovered, population-level analyses may shed light on which planets are able to evade runaway and which grow into gas giants.

TOI-1347 b also pushes the efficacy of photoevaporation to its limit. At $> 10~\Mearth$, the outflowing atmosphere has to overcome a deep gravitational potential well. On the other hand, the temperature of atmospheric outflows is likely capped below $10^4$~K due to strong radiative cooling at higher temperatures \citep[see, e.g.,][]{Murray-Clay}. An order of magnitude comparison reveals that the thermal sound speed ($\sim$10~{\kms} at $10^4$~K) may not overcome the escape velocity of the planet ($\sim$25~{\kms} at $10~\Mearth$ and $1.9~\Rearth$), preventing bulk hydrodynamic outflow (i.e. photoevaporation). Previous models showed that photoevaporation is significantly quenched on planets with heavier cores \citep[$\gtrsim 6~\Mearth$, see e.g.][]{OwenWu2017,WangDai2018}. Planets like TOI-1347~b are therefore important test cases to understand the limit of both photoevaporation and core-powered mass loss \citep{Ginzburg2018, Gupta2019}. Our mass and radius measurements disfavor the presence of a thick H/He envelope (Fig.~\ref{fig:mass-radius}). Did TOI-1347~b have, and then lose, a primordial H/He envelope? Or, could TOI-1347~b have formed near to its present-day scorching orbit without ever acquiring a substantial atmosphere? We encourage further investigation on this question.

\subsection{A Heavy-Mean-Molecular-Weight Atmosphere?}

Even though TOI-1347~b has a mass and radius that suggest an Earth-like bulk composition, we cannot rule out the presence of a heavy-mean-molecular-weight atmosphere. \citet{Lopez2017} showed that the largest of the non-giant USPs (such as 55~Cnc~e) can hold onto high-metallicity atmospheres even in the presence of strong stellar radiation. Such an atmosphere would only marginally inflate the planet's radius (the scale height of a CO$_2$ based atmosphere is about 11~km, while planetary radii are ~$\sim$10,000~km). 

In fact, our tentative phase curve (3$\sigma$) and secondary eclipse (2$\sigma$) detections of TOI-1347~b point to a nonzero albedo and a possible phase offset to the west. These features could indicate the presence of an atmosphere at least partially covered by reflective silicate clouds. It may be the case that the deep gravitational wells of the most massive super-Earth USPs are sufficient to cling to such an atmosphere and resist atmospheric loss mechanisms. Hu et al. (in prep.) show that their JWST NIRCam and MIRI observations of 55~Cnc~e, a similar massive USP (0.73-day orbit; 9$M_\oplus$), can only be explained if the planet still has a CO or CO$_2$ based atmosphere. Future observations with JWST might uncover a similar story for TOI-1347~b (TSM$=19$, ESM$=5.6$, using the equations of \citealt{Kempton2018}).

Alternatively, the high-amplitude \tess{} phase curve of TOI-1347~b may be a consequence of outgassed Na emission on the hot dayside of the planet. \citet{Zieba2022} showed that this emission can explain the phase curve of the lava world K2-141~b, which has been observed in both the \textit{Kepler} passband and \textit{Spitzer}-4.5~$\mu$m passband \citep{Malavolta2018}. In their analysis, they found that the two phase curves are inconsistent with a blackbody model, with the visible-light phase curve having a higher amplitude than expected. Similarly high visible-light phase curve amplitudes have been reported for the lava worlds 55~Cnc~e \citep{Kipping2020} and Kepler-10~b \citep{Batalha2011, Rouan2011}, although the latter has not yet been observed in the infrared, which would test blackbody emission. If Na emission is responsible for the observations of TOI-1347~b reported here, it may more easily explain the tentative phase curve offset than reflective clouds, which would need to be nonuniformly distributed across the planet.

\section{Summary}

We have characterized two transiting planets in the TOI-1347 system, TOI-1347~b, a USP (0.85~d), and its outer small companion TOI-1347~c (4.84~d). Using \tess{} photometry and an independent transit fitting pipeline, we measured a radius of $\Rb \pm \Rberr~\Rearth$ for the USP, and $\Rc \pm \Rcerr~\Rearth$ for its companion. We conducted a RV campaign of the TOI-1347 system with HIRES (as part of TKS) and with HARPS-N. We measured a mass of $\Mb \pm \Mberr~\Mearth$ for the USP, consistent with a bulk Earth-like composition and inconsistent with a H/He envelope (see Fig~\ref{fig:mass-radius}). This composition is perhaps unsurprising given the system age of $1.4 \pm 0.4$~Gyr, the short timescale on which intensive photoevaporation operates (few 100~Myr), and the high insolation flux at TOI 1347~b's orbit; any primordial H/He envelope should have been destroyed by now. We were unable to detect the companion TOI-1347~c with RVs. We placed a 95\% upper limit of $< \Mculim ~\Mearth$. Of note is the minimum mutual inclination between planets b and c implied by our measured orbital inclinations: $\sim 7^\circ$. This is unusually large compared to typical \textit{Kepler} multis \citep{Fabrycky2014}, which may be another indicator of the migration dynamics that produce USPs.

TOI-1347~b is the most massive of the $
< 2~\Rearth$ (i.e., primarily solid by volume) USPs to date. Its mass sets an upper limit on runaway accretion processes and places TOI-1347~b in a region of the mass-radius diagram in which the pressures and temperatures reached inside the planet have not been well characterized either by experiments or theoretical modeling.

Intriguingly, we measured a tentative (3$\sigma$) phase-curve variability, as well as a secondary eclipse (2$\sigma$) for the USP TOI-1347~b. The phase curve asymmetry strongly suggests an optically thick atmosphere. However, our mass and radius measurements of TOI-1347~b are highly inconsistent with any significant H/He envelope. As a result, any such atmosphere must have a high mean molecular weight. It could be comprised of reflective silicate clouds, or may be the result of the outgassing of Na from the molten surface. Future observations (e.g. with JWST) would help confirm such an atmosphere and reveal its composition.

\section{Acknowledgements}
    
Some of the data presented herein were obtained at the W. M. Keck Observatory, which is operated as a scientific partnership among the California Institute of Technology, the University of California, and the National Aeronautics and Space Administration. The Observatory was made possible by the generous financial support of the W. M. Keck Foundation. Keck Observatory occupies the summit of Maunakea, a place of significant ecological, cultural, and spiritual importance within the indigenous Hawaiian community. We understand and embrace our accountability to Maunakea and the indigenous Hawaiian community, and commit to our role in long-term mutual stewardship. We are most fortunate to have the opportunity to conduct observations from Maunakea.

This paper made use of data collected by the \tess{} mission and are publicly available from the Mikulski Archive for Space Telescopes (MAST) operated by the Space Telescope Science Institute (STScI). All the {\it TIC} data used in this paper can be found in MAST: \dataset[10.17909/fwdt-2x66]{http://dx.doi.org/10.17909/fwdt-2x66} \citep{ticdoi}. All the {\it TESS} data used in this paper can be found in MAST: \dataset[10.17909/t9-nmc8-f686]{http://dx.doi.org/t9-nmc8-f686} \citep{mastdoi}. Data from the NASA Exoplanet Archive can be found at \dataset[10.26133/NEA1]{http://dx.doi.org/10.26133/NEA1} \citep{neadoi}. Funding for the \tess{} mission is provided by NASA’s Science Mission Directorate. We acknowledge the use of public \tess{} data from pipelines at the \tess{} Science Office and at the \tess{} Science Processing Operations Center. This work is based also on observations made with the Italian Telescopio Nazionale Galileo (TNG) operated on the island of La Palma by the Fundaci\'on Galileo Galilei of the INAF (Istituto Nazionale di Astrofisica) at the Spanish Observatorio del Roque de los Muchachos of the Instituto de Astrofisica de Canarias.
Some of the observations in this paper made use of the High-Resolution Imaging instrument {\okina}Alopeke. {\okina}Alopeke was funded by the NASA Exoplanet Exploration Program and built at the NASA Ames Research Center by Steve B. Howell, Nic Scott, Elliott P. Horch, and Emmett Quigley. {\okina}Alopeke was mounted on the Gemini North telescope of the international Gemini Observatory, a program of NSF's NOIRLab, which is managed by the Association of Universities for Research in Astronomy (AURA) under a cooperative agreement with the National Science Foundation.

We thank the time assignment committees of the University of California, the California Institute of Technology, NASA, and the University of Hawaii for supporting the TESS-Keck Survey with observing time at Keck Observatory and on the Automated Planet Finder. We thank NASA for funding associated with our Key Strategic Mission Support project. We gratefully acknowledge the efforts and dedication of the Keck Observatory staff for support of HIRES and remote observing.

R.A.R. acknowledges support from the National Science Foundation through the Graduate Research Fellowship Program (DGE 1745301). M.R. acknowledges support from Heising-Simons grant \#2023-4478. D.H. acknowledges support from the Alfred P. Sloan Foundation, the National Aeronautics and Space Administration (80NSSC21K0652) and the Australian Research Council (FT200100871).

\software{
\texttt{astropy}    \citep{astropy},
\texttt{corner}     \citep{corner},
\texttt{exoplanet}  \citep{exoplanet},
\texttt{FASMA}      \citep{Tsantaki2020},
\texttt{lightkurve} \citep{lightkurve},
\texttt{matplotlib} \citep{matplotlib},
\texttt{numpy}      \citep{numpy},
\texttt{pandas}     \citep{pandas},
\texttt{radvel}     \citep{radvel},
\texttt{scipy}      \citep{scipy},
\texttt{SpecMatch-Synth} \citep{Petigura2015},
\texttt{spinspotter}\citep{spinspotter}
}

\facility{Keck-I: HIRES \citep{HIRES}, HARPS-N \citep{HARPSN}, \tess{} \citep{TESS}, Gemini North: {\okina}Alopeke \citep{Alopeke}
} 

\bibliography{references}

\begin{thebibliography}{}
\expandafter\ifx\csname natexlab\endcsname\relax\def\natexlab#1{#1}\fi
\providecommand{\url}[1]{\href{#1}{#1}}

\bibitem[{{Aigrain} \& {Foreman-Mackey}(2023)}]{Aigrain2023}
{Aigrain}, S., \& {Foreman-Mackey}, D. 2023, \araa, 61, 329

\bibitem[{{Armstrong} {et~al.}(2020){Armstrong}, {Lopez}, {Adibekyan}, {Booth},
  {Bryant}, {Collins}, {Deleuil}, {Emsenhuber}, {Huang}, {King}, {Lillo-Box},
  {Lissauer}, {Matthews}, {Mousis}, {Nielsen}, {Osborn}, {Otegi}, {Santos},
  {Sousa}, {Stassun}, {Veras}, {Ziegler}, {Acton}, {Almenara}, {Anderson},
  {Barrado}, {Barros}, {Bayliss}, {Belardi}, {Bouchy}, {Brice{\~n}o}, {Brogi},
  {Brown}, {Burleigh}, {Casewell}, {Chaushev}, {Ciardi}, {Collins},
  {Col{\'o}n}, {Cooke}, {Crossfield}, {D{\'\i}az}, {Delgado Mena}, {Demangeon},
  {Dorn}, {Dumusque}, {Eigm{\"u}ller}, {Fausnaugh}, {Figueira}, {Gan},
  {Gandhi}, {Gill}, {Gonzales}, {Goad}, {G{\"u}nther}, {Helled}, {Hojjatpanah},
  {Howell}, {Jackman}, {Jenkins}, {Jenkins}, {Jensen}, {Kennedy}, {Latham},
  {Law}, {Lendl}, {Lozovsky}, {Mann}, {Moyano}, {McCormac}, {Meru},
  {Mordasini}, {Osborn}, {Pollacco}, {Queloz}, {Raynard}, {Ricker}, {Rowden},
  {Santerne}, {Schlieder}, {Seager}, {Sha}, {Tan}, {Tilbrook}, {Ting}, {Udry},
  {Vanderspek}, {Watson}, {West}, {Wilson}, {Winn}, {Wheatley}, {Villasenor},
  {Vines}, \& {Zhan}}]{Armstrong849}
{Armstrong}, D.~J., {Lopez}, T.~A., {Adibekyan}, V., {et~al.} 2020, \nat, 583,
  39

\bibitem[{{Astropy Collaboration} {et~al.}(2022){Astropy Collaboration},
  {Price-Whelan}, {Lim}, {Earl}, {Starkman}, {Bradley}, {Shupe}, {Patil},
  {Corrales}, {Brasseur}, {N{\"o}the}, {Donath}, {Tollerud}, {Morris},
  {Ginsburg}, {Vaher}, {Weaver}, {Tocknell}, {Jamieson}, {van Kerkwijk},
  {Robitaille}, {Merry}, {Bachetti}, {G{\"u}nther}, {Aldcroft},
  {Alvarado-Montes}, {Archibald}, {B{\'o}di}, {Bapat}, {Barentsen},
  {Baz{\'a}n}, {Biswas}, {Boquien}, {Burke}, {Cara}, {Cara}, {Conroy},
  {Conseil}, {Craig}, {Cross}, {Cruz}, {D'Eugenio}, {Dencheva}, {Devillepoix},
  {Dietrich}, {Eigenbrot}, {Erben}, {Ferreira}, {Foreman-Mackey}, {Fox},
  {Freij}, {Garg}, {Geda}, {Glattly}, {Gondhalekar}, {Gordon}, {Grant},
  {Greenfield}, {Groener}, {Guest}, {Gurovich}, {Handberg}, {Hart},
  {Hatfield-Dodds}, {Homeier}, {Hosseinzadeh}, {Jenness}, {Jones}, {Joseph},
  {Kalmbach}, {Karamehmetoglu}, {Ka{\l}uszy{\'n}ski}, {Kelley}, {Kern},
  {Kerzendorf}, {Koch}, {Kulumani}, {Lee}, {Ly}, {Ma}, {MacBride}, {Maljaars},
  {Muna}, {Murphy}, {Norman}, {O'Steen}, {Oman}, {Pacifici}, {Pascual},
  {Pascual-Granado}, {Patil}, {Perren}, {Pickering}, {Rastogi}, {Roulston},
  {Ryan}, {Rykoff}, {Sabater}, {Sakurikar}, {Salgado}, {Sanghi}, {Saunders},
  {Savchenko}, {Schwardt}, {Seifert-Eckert}, {Shih}, {Jain}, {Shukla}, {Sick},
  {Simpson}, {Singanamalla}, {Singer}, {Singhal}, {Sinha}, {Sip{\H{o}}cz},
  {Spitler}, {Stansby}, {Streicher}, {{\v{S}}umak}, {Swinbank}, {Taranu},
  {Tewary}, {Tremblay}, {de Val-Borro}, {Van Kooten}, {Vasovi{\'c}}, {Verma},
  {de Miranda Cardoso}, {Williams}, {Wilson}, {Winkel}, {Wood-Vasey}, {Xue},
  {Yoachim}, {Zhang}, {Zonca}, \& {Astropy Project Contributors}}]{astropy}
{Astropy Collaboration}, {Price-Whelan}, A.~M., {Lim}, P.~L., {et~al.} 2022,
  \apj, 935, 167

\bibitem[{{Batalha} {et~al.}(2011){Batalha}, {Borucki}, {Bryson}, {Buchhave},
  {Caldwell}, {Christensen-Dalsgaard}, {Ciardi}, {Dunham}, {Fressin},
  {Gautier}, {Gilliland}, {Haas}, {Howell}, {Jenkins}, {Kjeldsen}, {Koch},
  {Latham}, {Lissauer}, {Marcy}, {Rowe}, {Sasselov}, {Seager}, {Steffen},
  {Torres}, {Basri}, {Brown}, {Charbonneau}, {Christiansen}, {Clarke},
  {Cochran}, {Dupree}, {Fabrycky}, {Fischer}, {Ford}, {Fortney}, {Girouard},
  {Holman}, {Johnson}, {Isaacson}, {Klaus}, {Machalek}, {Moorehead},
  {Morehead}, {Ragozzine}, {Tenenbaum}, {Twicken}, {Quinn}, {VanCleve},
  {Walkowicz}, {Welsh}, {Devore}, \& {Gould}}]{Batalha2011}
{Batalha}, N.~M., {Borucki}, W.~J., {Bryson}, S.~T., {et~al.} 2011, \apj, 729,
  27

\bibitem[{{Batygin} {et~al.}(2011){Batygin}, {Stevenson}, \&
  {Bodenheimer}}]{Batygin2011}
{Batygin}, K., {Stevenson}, D.~J., \& {Bodenheimer}, P.~H. 2011, \apj, 738, 1

\bibitem[{{Berger} {et~al.}(2018){Berger}, {Howard}, \&
  {Boesgaard}}]{Berger_li}
{Berger}, T.~A., {Howard}, A.~W., \& {Boesgaard}, A.~M. 2018, \apj, 855, 115

\bibitem[{{Berger} {et~al.}(2020){Berger}, {Huber}, {van Saders}, {Gaidos},
  {Tayar}, \& {Kraus}}]{Berger2020}
{Berger}, T.~A., {Huber}, D., {van Saders}, J.~L., {et~al.} 2020, \aj, 159, 280

\bibitem[{{Blunt} {et~al.}(2023){Blunt}, {Carvalho}, {David}, {Beichman},
  {Zink}, {Gaidos}, {Behmard}, {Bouma}, {Cody}, {Dai}, {Foreman-Mackey},
  {Grunblatt}, {Howard}, {Kosiarek}, {Knutson}, {Rubenzahl}, {Beard},
  {Chontos}, {Giacalone}, {Hirano}, {Johnson}, {Lubin}, {Akana Murphy},
  {Petigura}, {Van Zandt}, \& {Weiss}}]{Blunt2023}
{Blunt}, S., {Carvalho}, A., {David}, T.~J., {et~al.} 2023, \aj, 166, 62

\bibitem[{{Borucki} {et~al.}(2010){Borucki}, {Koch}, {Basri}, {Batalha},
  {Brown}, {Caldwell}, {Caldwell}, {Christensen-Dalsgaard}, {Cochran},
  {DeVore}, {Dunham}, {Dupree}, {Gautier}, {Geary}, {Gilliland}, {Gould},
  {Howell}, {Jenkins}, {Kondo}, {Latham}, {Marcy}, {Meibom}, {Kjeldsen},
  {Lissauer}, {Monet}, {Morrison}, {Sasselov}, {Tarter}, {Boss}, {Brownlee},
  {Owen}, {Buzasi}, {Charbonneau}, {Doyle}, {Fortney}, {Ford}, {Holman},
  {Seager}, {Steffen}, {Welsh}, {Rowe}, {Anderson}, {Buchhave}, {Ciardi},
  {Walkowicz}, {Sherry}, {Horch}, {Isaacson}, {Everett}, {Fischer}, {Torres},
  {Johnson}, {Endl}, {MacQueen}, {Bryson}, {Dotson}, {Haas}, {Kolodziejczak},
  {Van Cleve}, {Chandrasekaran}, {Twicken}, {Quintana}, {Clarke}, {Allen},
  {Li}, {Wu}, {Tenenbaum}, {Verner}, {Bruhweiler}, {Barnes}, \&
  {Prsa}}]{Borucki2010}
{Borucki}, W.~J., {Koch}, D., {Basri}, G., {et~al.} 2010, Science, 327, 977

\bibitem[{{Bouma} {et~al.}(2023){Bouma}, {Palumbo}, \&
  {Hillenbrand}}]{Bouma_2023}
{Bouma}, L.~G., {Palumbo}, E.~K., \& {Hillenbrand}, L.~A. 2023, \apjl, 947, L3

\bibitem[{{Butler} {et~al.}(1996){Butler}, {Marcy}, {Williams}, {McCarthy},
  {Dosanjh}, \& {Vogt}}]{Butler1996}
{Butler}, R.~P., {Marcy}, G.~W., {Williams}, E., {et~al.} 1996, \pasp, 108, 500

\bibitem[{{Chachan} {et~al.}(2021){Chachan}, {Lee}, \& {Knutson}}]{Chachan}
{Chachan}, Y., {Lee}, E.~J., \& {Knutson}, H.~A. 2021, arXiv e-prints,
  arXiv:2101.10333

\bibitem[{{Chen} \& {Rogers}(2016)}]{Chen_Rogers}
{Chen}, H., \& {Rogers}, L.~A. 2016, \apj, 831, 180

\bibitem[{{Chontos} {et~al.}(2022){Chontos}, {Murphy}, {MacDougall},
  {Fetherolf}, {Van Zandt}, {Rubenzahl}, {Beard}, {Huber}, {Batalha},
  {Crossfield}, {Dressing}, {Fulton}, {Howard}, {Isaacson}, {Kane}, {Petigura},
  {Robertson}, {Roy}, {Weiss}, {Behmard}, {Dai}, {Dalba}, {Giacalone}, {Hill},
  {Lubin}, {Mayo}, {Mo{\v{c}}nik}, {Polanski}, {Rosenthal}, {Scarsdale},
  {Turtelboom}, {Ricker}, {Vanderspek}, {Latham}, {Seager}, {Winn}, {Jenkins},
  {Quinn}, {Guerrero}, {Collins}, {Ciardi}, {Shporer}, {Goeke}, {Levine},
  {Ting}, {Bieryla}, {Collins}, {Kielkopf}, {Barkaoui}, {Benni},
  {Esparza-Borges}, {Conti}, {Hooton}, {Kagetani}, {Laloum}, {Marino},
  {Massey}, {Murgas}, {Papini}, {Schwarz}, {Srdoc}, {Stockdale}, {Wang},
  {Wittrock}, \& {Zou}}]{TKS0}
{Chontos}, A., {Murphy}, J. M.~A., {MacDougall}, M.~G., {et~al.} 2022, \aj,
  163, 297

\bibitem[{{Co{\c{s}}kuno{\v{g}}lu} {et~al.}(2011){Co{\c{s}}kuno{\v{g}}lu},
  {Ak}, {Bilir}, {Karaali}, {Yaz}, {Gilmore}, {Seabroke}, {Bienaym{\'e}},
  {Bland-Hawthorn}, {Campbell}, {Freeman}, {Gibson}, {Grebel}, {Munari},
  {Navarro}, {Parker}, {Siebert}, {Siviero}, {Steinmetz}, {Watson}, {Wyse}, \&
  {Zwitter}}]{LSR}
{Co{\c{s}}kuno{\v{g}}lu}, B., {Ak}, S., {Bilir}, S., {et~al.} 2011, \mnras,
  412, 1237

\bibitem[{Cosentino {et~al.}(2012)Cosentino, Lovis, Pepe, Cameron, Latham,
  Molinari, Udry, Bezawada, Black, Born, Buchschacher, Charbonneau, Figueira,
  Fleury, Galli, Gallie, Gao, Ghedina, Gonzalez, Gonzalez, Guerra, Henry,
  Horne, Hughes, Kelly, Lodi, Lunney, Maire, Mayor, Micela, Ordway, Peacock,
  Phillips, Piotto, Pollacco, Queloz, Rice, Riverol, Riverol, Juan, Sasselov,
  Segransan, Sozzetti, Sosnowska, Stobie, Szentgyorgyi, Vick, \&
  Weber}]{HARPSN}
Cosentino, R., Lovis, C., Pepe, F., {et~al.} 2012, in Ground-based and Airborne
  Instrumentation for Astronomy IV, ed. I.~S. McLean, S.~K. Ramsay, \&
  H.~Takami, Vol. 8446, International Society for Optics and Photonics (SPIE),
  84461V.
\newblock \url{https://doi.org/10.1117/12.925738}

\bibitem[{{Czesla} {et~al.}(2019){Czesla}, {Schr{\"o}ter}, {Schneider},
  {Huber}, {Pfeifer}, {Andreasen}, \& {Zechmeister}}]{PyAstronomy}
{Czesla}, S., {Schr{\"o}ter}, S., {Schneider}, C.~P., {et~al.} 2019, {PyA:
  Python astronomy-related packages}, , , ascl:1906.010

\bibitem[{{Dai} {et~al.}(2019){Dai}, {Masuda}, {Winn}, \& {Zeng}}]{Dai2019}
{Dai}, F., {Masuda}, K., {Winn}, J.~N., \& {Zeng}, L. 2019, \apj, 883, 79

\bibitem[{{Dai} {et~al.}(2021){Dai}, {Howard}, {Batalha}, {Beard}, {Behmard},
  {Blunt}, {Brinkman}, {Chontos}, {Crossfield}, {Dalba}, {Dressing}, {Fulton},
  {Giacalone}, {Hill}, {Huber}, {Isaacson}, {Kane}, {Lubin}, {Mayo},
  {Mo{\v{c}}nik}, {Akana Murphy}, {Petigura}, {Rice}, {Robertson}, {Rosenthal},
  {Roy}, {Rubenzahl}, {Weiss}, {Zandt}, {Beichman}, {Ciardi}, {Collins},
  {Gonzales}, {Howell}, {Matson}, {Matthews}, {Schlieder}, {Schwarz}, {Ricker},
  {Vanderspek}, {Latham}, {Seager}, {Winn}, {Jenkins}, {Caldwell}, {Colon},
  {Dragomir}, {Lund}, {McLean}, {Rudat}, \& {Shporer}}]{Dai_1444}
{Dai}, F., {Howard}, A.~W., {Batalha}, N.~M., {et~al.} 2021, \aj, 162, 62

\bibitem[{{Dawson} \& {Fabrycky}(2010)}]{DawsonFabrycky2010}
{Dawson}, R.~I., \& {Fabrycky}, D.~C. 2010, \apj, 722, 937

\bibitem[{{Demory} {et~al.}(2013){Demory}, {de Wit}, {Lewis}, {Fortney},
  {Zsom}, {Seager}, {Knutson}, {Heng}, {Madhusudhan}, {Gillon}, {Barclay},
  {Desert}, {Parmentier}, \& {Cowan}}]{Demory2013}
{Demory}, B.-O., {de Wit}, J., {Lewis}, N., {et~al.} 2013, \apjl, 776, L25

\bibitem[{{Demory} {et~al.}(2016){Demory}, {Gillon}, {de Wit}, {Madhusudhan},
  {Bolmont}, {Heng}, {Kataria}, {Lewis}, {Hu}, {Krick}, {Stamenkovi{\'c}},
  {Benneke}, {Kane}, \& {Queloz}}]{Demory55cnce_phasecurve}
{Demory}, B.-O., {Gillon}, M., {de Wit}, J., {et~al.} 2016, \nat, 532, 207

\bibitem[{{Duane} {et~al.}(1987){Duane}, {Kennedy}, {Pendleton}, \&
  {Roweth}}]{hmc:duane87}
{Duane}, S., {Kennedy}, A.~D., {Pendleton}, B.~J., \& {Roweth}, D. 1987,
  Physics Letters B, 195, 216

\bibitem[{{Dumusque} {et~al.}(2021){Dumusque}, {Cretignier}, {Sosnowska},
  {Buchschacher}, {Lovis}, {Phillips}, {Pepe}, {Alesina}, {Buchhave},
  {Burnier}, {Cecconi}, {Cegla}, {Cloutier}, {Collier Cameron}, {Cosentino},
  {Ghedina}, {Gonz{\'a}lez}, {Haywood}, {Latham}, {Lodi}, {L{\'o}pez-Morales},
  {Maldonado}, {Malavolta}, {Micela}, {Molinari}, {Mortier}, {P{\'e}rez
  Ventura}, {Pinamonti}, {Poretti}, {Rice}, {Riverol}, {Riverol}, {San Juan},
  {S{\'e}gransan}, {Sozzetti}, {Thompson}, {Udry}, \& {Wilson}}]{Dumusque2021}
{Dumusque}, X., {Cretignier}, M., {Sosnowska}, D., {et~al.} 2021, \aap, 648,
  A103

\bibitem[{{Duncan} {et~al.}(1991){Duncan}, {Vaughan}, {Wilson}, {Preston},
  {Frazer}, {Lanning}, {Misch}, {Mueller}, {Soyumer}, {Woodard}, {Baliunas},
  {Noyes}, {Hartmann}, {Porter}, {Zwaan}, {Middelkoop}, {Rutten}, \&
  {Mihalas}}]{Sindex}
{Duncan}, D.~K., {Vaughan}, A.~H., {Wilson}, O.~C., {et~al.} 1991, \apjs, 76,
  383

\bibitem[{{Essack} {et~al.}(2023){Essack}, {Shporer}, {Burt}, {Seager},
  {Cambioni}, {Lin}, {Collins}, {Mamajek}, {Stassun}, {Ricker}, {Vanderspek},
  {Latham}, {Winn}, {Jenkins}, {Butler}, {Charbonneau}, {Collins}, {Crane},
  {Gan}, {Hellier}, {Howell}, {Irwin}, {Mann}, {Ramadhan}, {Shectman}, {Teske},
  {Yee}, {Mireles}, {Quintana}, {Tenenbaum}, {Torres}, \& {Furlan}}]{toi1075}
{Essack}, Z., {Shporer}, A., {Burt}, J.~A., {et~al.} 2023, \aj, 165, 47

\bibitem[{{Fabrycky} {et~al.}(2014){Fabrycky}, {Lissauer}, {Ragozzine}, {Rowe},
  {Steffen}, {Agol}, {Barclay}, {Batalha}, {Borucki}, {Ciardi}, {Ford},
  {Gautier}, {Geary}, {Holman}, {Jenkins}, {Li}, {Morehead}, {Morris},
  {Shporer}, {Smith}, {Still}, \& {Van Cleve}}]{Fabrycky2014}
{Fabrycky}, D.~C., {Lissauer}, J.~J., {Ragozzine}, D., {et~al.} 2014, \apj,
  790, 146

\bibitem[{{Foreman-Mackey}(2016)}]{corner}
{Foreman-Mackey}, D. 2016, JOSS, 1

\bibitem[{Foreman-Mackey {et~al.}(2017)Foreman-Mackey, Agol, Ambikasaran, \&
  Angus}]{celerite}
Foreman-Mackey, D., Agol, E., Ambikasaran, S., \& Angus, R. 2017, The
  Astronomical Journal, 154, 220.
\newblock \url{https://dx.doi.org/10.3847/1538-3881/aa9332}

\bibitem[{Foreman-Mackey {et~al.}(2013)Foreman-Mackey, Hogg, Lang, \&
  Goodman}]{emcee}
Foreman-Mackey, D., Hogg, D.~W., Lang, D., \& Goodman, J. 2013, \pasp, 125, 306

\bibitem[{{Foreman-Mackey} {et~al.}(2021){Foreman-Mackey}, {Luger}, {Agol},
  {Barclay}, {Bouma}, {Brandt}, {Czekala}, {David}, {Dong}, {Gilbert},
  {Gordon}, {Hedges}, {Hey}, {Morris}, {Price-Whelan}, \& {Savel}}]{exoplanet}
{Foreman-Mackey}, D., {Luger}, R., {Agol}, E., {et~al.} 2021, arXiv e-prints,
  arXiv:2105.01994

\bibitem[{{Fulton} \& {Petigura}(2018)}]{FGap2018}
{Fulton}, B.~J., \& {Petigura}, E.~A. 2018, ArXiv e-prints, arXiv:1805.01453

\bibitem[{{Fulton} {et~al.}(2018){Fulton}, {Petigura}, {Blunt}, \&
  {Sinukoff}}]{radvel}
{Fulton}, B.~J., {Petigura}, E.~A., {Blunt}, S., \& {Sinukoff}, E. 2018, \pasp,
  130, 044504

\bibitem[{{Fulton} {et~al.}(2017){Fulton}, {Petigura}, {Howard}, {Isaacson},
  {Marcy}, {Cargile}, {Hebb}, {Weiss}, {Johnson}, {Morton}, {Sinukoff},
  {Crossfield}, \& {Hirsch}}]{FGap2017}
{Fulton}, B.~J., {Petigura}, E.~A., {Howard}, A.~W., {et~al.} 2017, \aj, 154,
  109

\bibitem[{{Gaia Collaboration} {et~al.}(2023){Gaia Collaboration}, {Vallenari},
  {Brown}, {Prusti}, {de Bruijne}, {Arenou}, {Babusiaux}, {Biermann},
  {Creevey}, {Ducourant}, {Evans}, {Eyer}, {Guerra}, {Hutton}, {Jordi},
  {Klioner}, {Lammers}, {Lindegren}, {Luri}, {Mignard}, {Panem}, {Pourbaix},
  {Randich}, {Sartoretti}, {Soubiran}, {Tanga}, {Walton}, {Bailer-Jones},
  {Bastian}, {Drimmel}, {Jansen}, {Katz}, {Lattanzi}, {van Leeuwen}, {Bakker},
  {Cacciari}, {Casta{\~n}eda}, {De Angeli}, {Fabricius}, {Fouesneau},
  {Fr{\'e}mat}, {Galluccio}, {Guerrier}, {Heiter}, {Masana}, {Messineo},
  {Mowlavi}, {Nicolas}, {Nienartowicz}, {Pailler}, {Panuzzo}, {Riclet}, {Roux},
  {Seabroke}, {Sordo}, {Th{\'e}venin}, {Gracia-Abril}, {Portell}, {Teyssier},
  {Altmann}, {Andrae}, {Audard}, {Bellas-Velidis}, {Benson}, {Berthier},
  {Blomme}, {Burgess}, {Busonero}, {Busso}, {C{\'a}novas}, {Carry}, {Cellino},
  {Cheek}, {Clementini}, {Damerdji}, {Davidson}, {de Teodoro}, {Nu{\~n}ez
  Campos}, {Delchambre}, {Dell'Oro}, {Esquej}, {Fern{\'a}ndez-Hern{\'a}ndez},
  {Fraile}, {Garabato}, {Garc{\'\i}a-Lario}, {Gosset}, {Haigron}, {Halbwachs},
  {Hambly}, {Harrison}, {Hern{\'a}ndez}, {Hestroffer}, {Hodgkin}, {Holl},
  {Jan{\ss}en}, {Jevardat de Fombelle}, {Jordan}, {Krone-Martins}, {Lanzafame},
  {L{\"o}ffler}, {Marchal}, {Marrese}, {Moitinho}, {Muinonen}, {Osborne},
  {Pancino}, {Pauwels}, {Recio-Blanco}, {Reyl{\'e}}, {Riello}, {Rimoldini},
  {Roegiers}, {Rybizki}, {Sarro}, {Siopis}, {Smith}, {Sozzetti}, {Utrilla},
  {van Leeuwen}, {Abbas}, {{\'A}brah{\'a}m}, {Abreu Aramburu}, {Aerts},
  {Aguado}, {Ajaj}, {Aldea-Montero}, {Altavilla}, {{\'A}lvarez}, {Alves},
  {Anders}, {Anderson}, {Anglada Varela}, {Antoja}, {Baines}, {Baker},
  {Balaguer-N{\'u}{\~n}ez}, {Balbinot}, {Balog}, {Barache}, {Barbato},
  {Barros}, {Barstow}, {Bartolom{\'e}}, {Bassilana}, {Bauchet}, {Becciani},
  {Bellazzini}, {Berihuete}, {Bernet}, {Bertone}, {Bianchi}, {Binnenfeld},
  {Blanco-Cuaresma}, {Blazere}, {Boch}, {Bombrun}, {Bossini}, {Bouquillon},
  {Bragaglia}, {Bramante}, {Breedt}, {Bressan}, {Brouillet}, {Brugaletta},
  {Bucciarelli}, {Burlacu}, {Butkevich}, {Buzzi}, {Caffau}, {Cancelliere},
  {Cantat-Gaudin}, {Carballo}, {Carlucci}, {Carnerero}, {Carrasco},
  {Casamiquela}, {Castellani}, {Castro-Ginard}, {Chaoul}, {Charlot}, {Chemin},
  {Chiaramida}, {Chiavassa}, {Chornay}, {Comoretto}, {Contursi}, {Cooper},
  {Cornez}, {Cowell}, {Crifo}, {Cropper}, {Crosta}, {Crowley}, {Dafonte},
  {Dapergolas}, {David}, {David}, {de Laverny}, {De Luise}, {De March}, {De
  Ridder}, {de Souza}, {de Torres}, {del Peloso}, {del Pozo}, {Delbo},
  {Delgado}, {Delisle}, {Demouchy}, {Dharmawardena}, {Di Matteo}, {Diakite},
  {Diener}, {Distefano}, {Dolding}, {Edvardsson}, {Enke}, {Fabre}, {Fabrizio},
  {Faigler}, {Fedorets}, {Fernique}, {Fienga}, {Figueras}, {Fournier},
  {Fouron}, {Fragkoudi}, {Gai}, {Garcia-Gutierrez}, {Garcia-Reinaldos},
  {Garc{\'\i}a-Torres}, {Garofalo}, {Gavel}, {Gavras}, {Gerlach}, {Geyer},
  {Giacobbe}, {Gilmore}, {Girona}, {Giuffrida}, {Gomel}, {Gomez},
  {Gonz{\'a}lez-N{\'u}{\~n}ez}, {Gonz{\'a}lez-Santamar{\'\i}a},
  {Gonz{\'a}lez-Vidal}, {Granvik}, {Guillout}, {Guiraud},
  {Guti{\'e}rrez-S{\'a}nchez}, {Guy}, {Hatzidimitriou}, {Hauser}, {Haywood},
  {Helmer}, {Helmi}, {Sarmiento}, {Hidalgo}, {Hilger}, {H{\l}adczuk}, {Hobbs},
  {Holland}, {Huckle}, {Jardine}, {Jasniewicz}, {Jean-Antoine Piccolo},
  {Jim{\'e}nez-Arranz}, {Jorissen}, {Juaristi Campillo}, {Julbe}, {Karbevska},
  {Kervella}, {Khanna}, {Kontizas}, {Kordopatis}, {Korn}, {K{\'o}sp{\'a}l},
  {Kostrzewa-Rutkowska}, {Kruszy{\'n}ska}, {Kun}, {Laizeau}, {Lambert},
  {Lanza}, {Lasne}, {Le Campion}, {Lebreton}, {Lebzelter}, {Leccia}, {Leclerc},
  {Lecoeur-Taibi}, {Liao}, {Licata}, {Lindstr{\o}m}, {Lister}, {Livanou},
  {Lobel}, {Lorca}, {Loup}, {Madrero Pardo}, {Magdaleno Romeo}, {Managau},
  {Mann}, {Manteiga}, {Marchant}, {Marconi}, {Marcos}, {Marcos Santos},
  {Mar{\'\i}n Pina}, {Marinoni}, {Marocco}, {Marshall}, {Martin Polo},
  {Mart{\'\i}n-Fleitas}, {Marton}, {Mary}, {Masip}, {Massari},
  {Mastrobuono-Battisti}, {Mazeh}, {McMillan}, {Messina}, {Michalik}, {Millar},
  {Mints}, {Molina}, {Molinaro}, {Moln{\'a}r}, {Monari}, {Mongui{\'o}},
  {Montegriffo}, {Montero}, {Mor}, {Mora}, {Morbidelli}, {Morel}, {Morris},
  {Muraveva}, {Murphy}, {Musella}, {Nagy}, {Noval}, {Oca{\~n}a}, {Ogden},
  {Ordenovic}, {Osinde}, {Pagani}, {Pagano}, {Palaversa}, {Palicio},
  {Pallas-Quintela}, {Panahi}, {Payne-Wardenaar}, {Pe{\~n}alosa Esteller},
  {Penttil{\"a}}, {Pichon}, {Piersimoni}, {Pineau}, {Plachy}, {Plum}, {Poggio},
  {Pr{\v{s}}a}, {Pulone}, {Racero}, {Ragaini}, {Rainer}, {Raiteri}, {Rambaux},
  {Ramos}, {Ramos-Lerate}, {Re Fiorentin}, {Regibo}, {Richards}, {Rios Diaz},
  {Ripepi}, {Riva}, {Rix}, {Rixon}, {Robichon}, {Robin}, {Robin}, {Roelens},
  {Rogues}, {Rohrbasser}, {Romero-G{\'o}mez}, {Rowell}, {Royer}, {Ruz Mieres},
  {Rybicki}, {Sadowski}, {S{\'a}ez N{\'u}{\~n}ez}, {Sagrist{\`a} Sell{\'e}s},
  {Sahlmann}, {Salguero}, {Samaras}, {Sanchez Gimenez}, {Sanna},
  {Santove{\~n}a}, {Sarasso}, {Schultheis}, {Sciacca}, {Segol}, {Segovia},
  {S{\'e}gransan}, {Semeux}, {Shahaf}, {Siddiqui}, {Siebert}, {Siltala},
  {Silvelo}, {Slezak}, {Slezak}, {Smart}, {Snaith}, {Solano}, {Solitro},
  {Souami}, {Souchay}, {Spagna}, {Spina}, {Spoto}, {Steele},
  {Steidelm{\"u}ller}, {Stephenson}, {S{\"u}veges}, {Surdej}, {Szabados},
  {Szegedi-Elek}, {Taris}, {Taylor}, {Teixeira}, {Tolomei}, {Tonello}, {Torra},
  {Torra}, {Torralba Elipe}, {Trabucchi}, {Tsounis}, {Turon}, {Ulla}, {Unger},
  {Vaillant}, {van Dillen}, {van Reeven}, {Vanel}, {Vecchiato}, {Viala},
  {Vicente}, {Voutsinas}, {Weiler}, {Wevers}, {Wyrzykowski}, {Yoldas}, {Yvard},
  {Zhao}, {Zorec}, {Zucker}, \& {Zwitter}}]{gaiadr3}
{Gaia Collaboration}, {Vallenari}, A., {Brown}, A.~G.~A., {et~al.} 2023, \aap,
  674, A1

\bibitem[{{Giles} {et~al.}(2017){Giles}, {Collier Cameron}, \&
  {Haywood}}]{Giles2017}
{Giles}, H. A.~C., {Collier Cameron}, A., \& {Haywood}, R.~D. 2017, \mnras,
  472, 1618

\bibitem[{Ginzburg {et~al.}(2018)Ginzburg, Schlichting, \& Sari}]{Ginzburg2018}
Ginzburg, S., Schlichting, H.~E., \& Sari, R. 2018, Monthly Notices of the
  Royal Astronomical Society, 476, 759.
\newblock \url{https://doi.org/10.1093/mnras/sty290}

\bibitem[{Grunblatt {et~al.}(2017)Grunblatt, Huber, Gaidos, Lopez, Howard,
  Isaacson, Sinukoff, Vanderburg, Nofi, Yu, North, Chaplin, Foreman-Mackey,
  Petigura, Ansdell, Weiss, Fulton, \& Lin}]{Grunblatt2017}
Grunblatt, S.~K., Huber, D., Gaidos, E., {et~al.} 2017, The Astronomical
  Journal, 154, 254.
\newblock \url{https://dx.doi.org/10.3847/1538-3881/aa932d}

\bibitem[{{Guerrero} {et~al.}(2021){Guerrero}, {Seager}, {Huang}, {Vanderburg},
  {Garcia Soto}, {Mireles}, {Hesse}, {Fong}, {Glidden}, {Shporer}, {Latham},
  {Collins}, {Quinn}, {Burt}, {Dragomir}, {Crossfield}, {Vanderspek},
  {Fausnaugh}, {Burke}, {Ricker}, {Daylan}, {Essack}, {G{\"u}nther}, {Osborn},
  {Pepper}, {Rowden}, {Sha}, {Villanueva}, {Yahalomi}, {Yu}, {Ballard},
  {Batalha}, {Berardo}, {Chontos}, {Dittmann}, {Esquerdo}, {Mikal-Evans},
  {Jayaraman}, {Krishnamurthy}, {Louie}, {Mehrle}, {Niraula}, {Rackham},
  {Rodriguez}, {Rowden}, {Sousa-Silva}, {Watanabe}, {Wong}, {Zhan},
  {Zivanovic}, {Christiansen}, {Ciardi}, {Swain}, {Lund}, {Mullally},
  {Fleming}, {Rodriguez}, {Boyd}, {Quintana}, {Barclay}, {Col{\'o}n},
  {Rinehart}, {Schlieder}, {Clampin}, {Jenkins}, {Twicken}, {Caldwell},
  {Coughlin}, {Henze}, {Lissauer}, {Morris}, {Rose}, {Smith}, {Tenenbaum},
  {Ting}, {Wohler}, {Bakos}, {Bean}, {Berta-Thompson}, {Bieryla}, {Bouma},
  {Buchhave}, {Butler}, {Charbonneau}, {Doty}, {Ge}, {Holman}, {Howard},
  {Kaltenegger}, {Kane}, {Kjeldsen}, {Kreidberg}, {Lin}, {Minsky}, {Narita},
  {Paegert}, {P{\'a}l}, {Palle}, {Sasselov}, {Spencer}, {Sozzetti}, {Stassun},
  {Torres}, {Udry}, \& {Winn}}]{Guerrero}
{Guerrero}, N.~M., {Seager}, S., {Huang}, C.~X., {et~al.} 2021, \apjs, 254, 39

\bibitem[{Gupta \& Schlichting(2019)}]{Gupta2019}
Gupta, A., \& Schlichting, H.~E. 2019, Monthly Notices of the Royal
  Astronomical Society, 487, 24.
\newblock \url{https://doi.org/10.1093/mnras/stz1230}

\bibitem[{Harris {et~al.}(2020)Harris, Millman, van~der Walt, Gommers,
  Virtanen, Cournapeau, Wieser, Taylor, Berg, Smith, Kern, Picus, Hoyer, van
  Kerkwijk, Brett, Haldane, Fernández~del Río, Wiebe, Peterson,
  Gérard-Marchant, Sheppard, Reddy, Weckesser, Abbasi, Gohlke, \&
  Oliphant}]{numpy}
Harris, C.~R., Millman, K.~J., van~der Walt, S.~J., {et~al.} 2020, \nat, 585,
  357–362

\bibitem[{{Hastings}(1970)}]{hastings70}
{Hastings}, W.~K. 1970, Biometrika, 57, 97

\bibitem[{Haywood {et~al.}(2014)Haywood, Collier~Cameron, Queloz, Barros,
  Deleuil, Fares, Gillon, Lanza, Lovis, Moutou, Pepe, Pollacco, Santerne,
  Ségransan, \& Unruh}]{Haywood2014}
Haywood, R.~D., Collier~Cameron, A., Queloz, D., {et~al.} 2014, Monthly Notices
  of the Royal Astronomical Society, 443, 2517.
\newblock \url{https://doi.org/10.1093/mnras/stu1320}

\bibitem[{Haywood {et~al.}(2016)Haywood, Collier~Cameron, Unruh, Lovis, Lanza,
  Llama, Deleuil, Fares, Gillon, Moutou, Pepe, Pollacco, Queloz, \&
  Ségransan}]{Haywood2016}
Haywood, R.~D., Collier~Cameron, A., Unruh, Y.~C., {et~al.} 2016, Monthly
  Notices of the Royal Astronomical Society, 457, 3637.
\newblock \url{https://doi.org/10.1093/mnras/stw187}

\bibitem[{Hedges {et~al.}(2020)Hedges, Angus, Barentsen, Saunders, Montet, \&
  Gully-Santiago}]{TESS_SIP}
Hedges, C., Angus, R., Barentsen, G., {et~al.} 2020, Research Notes of the AAS,
  4, 220.
\newblock \url{https://dx.doi.org/10.3847/2515-5172/abd106}

\bibitem[{Hoffman \& Gelman(2014)}]{nuts:hoffman14}
Hoffman, M., \& Gelman, A. 2014, J. Mach. Learn. Res., 15, 1593

\bibitem[{Holcomb {et~al.}(2022)Holcomb, Robertson, Hartigan, Oelkers, \&
  Robinson}]{spinspotter}
Holcomb, R.~J., Robertson, P., Hartigan, P., Oelkers, R.~J., \& Robinson, C.
  2022, The Astrophysical Journal, 936, 138.
\newblock \url{https://dx.doi.org/10.3847/1538-4357/ac8990}

\bibitem[{Howard {et~al.}(2010)Howard, Johnson, Marcy, Fischer, Wright, Bernat,
  Henry, Peek, Isaacson, Apps, Endl, Cochran, Valenti, Anderson, \&
  Piskunov}]{Howard2010}
Howard, A.~W., Johnson, J.~A., Marcy, G.~W., {et~al.} 2010, \apj, 721, 1467

\bibitem[{{Howell} {et~al.}(2011){Howell}, {Everett}, {Sherry}, {Horch}, \&
  {Ciardi}}]{howell:2011}
{Howell}, S.~B., {Everett}, M.~E., {Sherry}, W., {Horch}, E., \& {Ciardi},
  D.~R. 2011, \aj, 142, 19

\bibitem[{{Huber} {et~al.}(2017){Huber}, {Zinn}, {Bojsen-Hansen},
  {Pinsonneault}, {Sahlholdt}, {Serenelli}, {Silva Aguirre}, {Stassun},
  {Stello}, {Tayar}, {Bastien}, {Bedding}, {Buchhave}, {Chaplin}, {Davies},
  {Garc{\'\i}a}, {Latham}, {Mathur}, {Mosser}, \& {Sharma}}]{Huber17}
{Huber}, D., {Zinn}, J., {Bojsen-Hansen}, M., {et~al.} 2017, \apj, 844, 102

\bibitem[{{Hunter}(2007)}]{matplotlib}
{Hunter}, J.~D. 2007, CSE, 9, 90

\bibitem[{{Isaacson} \& {Fischer}(2010)}]{Isaacson}
{Isaacson}, H., \& {Fischer}, D. 2010, \apj, 725, 875

\bibitem[{{Jenkins} {et~al.}(2016){Jenkins}, {Twicken}, {McCauliff},
  {Campbell}, {Sanderfer}, {Lung}, {Mansouri-Samani}, {Girouard}, {Tenenbaum},
  {Klaus}, {Smith}, {Caldwell}, {Chacon}, {Henze}, {Heiges}, {Latham},
  {Morgan}, {Swade}, {Rinehart}, \& {Vanderspek}}]{jenkinsSPOC2016}
{Jenkins}, J.~M., {Twicken}, J.~D., {McCauliff}, S., {et~al.} 2016, in
  \procspie, Vol. 9913, Software and Cyberinfrastructure for Astronomy IV,
  99133E

\bibitem[{{Jenkins} {et~al.}(2020){Jenkins}, {D{\'\i}az}, {Kurtovic},
  {Espinoza}, {Vines}, {Rojas}, {Brahm}, {Torres}, {Cort{\'e}s-Zuleta}, {Soto},
  {Lopez}, {King}, {Wheatley}, {Winn}, {Ciardi}, {Ricker}, {Vanderspek},
  {Latham}, {Seager}, {Jenkins}, {Beichman}, {Bieryla}, {Burke},
  {Christiansen}, {Henze}, {Klaus}, {McCauliff}, {Mori}, {Narita}, {Nishiumi},
  {Tamura}, {de Leon}, {Quinn}, {Villase{\~n}or}, {Vezie}, {Lissauer},
  {Collins}, {Collins}, {Isopi}, {Mallia}, {Ercolino}, {Petrovich},
  {Jord{\'a}n}, {Acton}, {Armstrong}, {Bayliss}, {Bouchy}, {Belardi}, {Bryant},
  {Burleigh}, {Cabrera}, {Casewell}, {Chaushev}, {Cooke}, {Eigm{\"u}ller},
  {Erikson}, {Foxell}, {G{\"a}nsicke}, {Gill}, {Gillen}, {G{\"u}nther}, {Goad},
  {Hooton}, {Jackman}, {Louden}, {McCormac}, {Moyano}, {Nielsen}, {Pollacco},
  {Queloz}, {Rauer}, {Raynard}, {Smith}, {Tilbrook}, {Titz-Weider}, {Turner},
  {Udry}, {Walker}, {Watson}, {West}, {Palle}, {Ziegler}, {Law}, \&
  {Mann}}]{Jenkins}
{Jenkins}, J.~S., {D{\'\i}az}, M.~R., {Kurtovic}, N.~T., {et~al.} 2020, Nature
  Astronomy, 4, 1148

\bibitem[{{Kane} {et~al.}(2020){Kane}, {Roettenbacher}, {Unterborn}, {Foley},
  \& {Hill}}]{Kane2020}
{Kane}, S.~R., {Roettenbacher}, R.~M., {Unterborn}, C.~T., {Foley}, B.~J., \&
  {Hill}, M.~L. 2020, \psj, 1, 36

\bibitem[{{Kempton} {et~al.}(2018){Kempton}, {Bean}, {Louie}, {Deming}, {Koll},
  {Mansfield}, {Christiansen}, {L{\'o}pez-Morales}, {Swain}, {Zellem},
  {Ballard}, {Barclay}, {Barstow}, {Batalha}, {Beatty}, {Berta-Thompson},
  {Birkby}, {Buchhave}, {Charbonneau}, {Cowan}, {Crossfield}, {de Val-Borro},
  {Doyon}, {Dragomir}, {Gaidos}, {Heng}, {Hu}, {Kane}, {Kreidberg}, {Mallonn},
  {Morley}, {Narita}, {Nascimbeni}, {Pall{\'e}}, {Quintana}, {Rauscher},
  {Seager}, {Shkolnik}, {Sing}, {Sozzetti}, {Stassun}, {Valenti}, \& {von
  Essen}}]{Kempton2018}
{Kempton}, E. M.~R., {Bean}, J.~L., {Louie}, D.~R., {et~al.} 2018, \pasp, 130,
  114401

\bibitem[{{Kipping} \& {Jansen}(2020)}]{Kipping2020}
{Kipping}, D., \& {Jansen}, T. 2020, Research Notes of the American
  Astronomical Society, 4, 170

\bibitem[{{Kipping}(2013)}]{Kipping}
{Kipping}, D.~M. 2013, \mnras, 435, 2152

\bibitem[{{Kov{\'a}cs} {et~al.}(2002){Kov{\'a}cs}, {Zucker}, \&
  {Mazeh}}]{Kovac2002}
{Kov{\'a}cs}, G., {Zucker}, S., \& {Mazeh}, T. 2002, \aap, 391, 369

\bibitem[{{Kreidberg}(2015)}]{Kreidberg2015}
{Kreidberg}, L. 2015, \pasp, 127, 1161

\bibitem[{{Kreidberg} {et~al.}(2019){Kreidberg}, {Koll}, {Morley}, {Hu},
  {Schaefer}, {Deming}, {Stevenson}, {Dittmann}, {Vanderburg}, {Berardo},
  {Guo}, {Stassun}, {Crossfield}, {Charbonneau}, {Latham}, {Loeb}, {Ricker},
  {Seager}, \& {Vanderspek}}]{LHS3844}
{Kreidberg}, L., {Koll}, D. D.~B., {Morley}, C., {et~al.} 2019, \nat, 573, 87

\bibitem[{{Lee}(2019)}]{Lee_accretion}
{Lee}, E.~J. 2019, \apj, 878, 36

\bibitem[{{Lightkurve Collaboration} {et~al.}(2018){Lightkurve Collaboration},
  {Cardoso}, {Hedges}, {Gully-Santiago}, {Saunders}, {Cody}, {Barclay}, {Hall},
  {Sagear}, {Turtelboom}, {Zhang}, {Tzanidakis}, {Mighell}, {Coughlin}, {Bell},
  {Berta-Thompson}, {Williams}, {Dotson}, \& {Barentsen}}]{lightkurve}
{Lightkurve Collaboration}, {Cardoso}, J.~V.~d.~M., {Hedges}, C., {et~al.}
  2018, {Lightkurve: Kepler and TESS time series analysis in Python},
  Astrophysics Source Code Library, , , ascl:1812.013

\bibitem[{{Lopez}(2017)}]{Lopez2017}
{Lopez}, E.~D. 2017, \mnras, 472, 245

\bibitem[{MacDougall {et~al.}(2023)MacDougall, Petigura, Gilbert, Angelo,
  Batalha, Beard, Behmard, Blunt, Brinkman, Chontos, Crossfield, Dai, Dalba,
  Dressing, Fetherolf, Fulton, Giacalone, Hill, Holcomb, Howard, Huber,
  Isaacson, Kane, Kosiarek, Lubin, Mayo, Močnik, Murphy, Pidhorodetska,
  Polanski, Rice, Robertson, Rosenthal, Roy, Rubenzahl, Scarsdale, Turtelboom,
  Tyler, Zandt, Weiss, \& Yee}]{MacDougall2023}
MacDougall, M.~G., Petigura, E.~A., Gilbert, G.~J., {et~al.} 2023, The
  Astronomical Journal, 166, 33.
\newblock \url{https://dx.doi.org/10.3847/1538-3881/acd557}

\bibitem[{{Malavolta} {et~al.}(2018){Malavolta}, {Mayo}, {Louden}, {Rajpaul},
  {Bonomo}, {Buchhave}, {Kreidberg}, {Kristiansen}, {Lopez-Morales}, {Mortier},
  {Vand erburg}, {Coffinet}, {Ehrenreich}, {Lovis}, {Bouchy}, {Charbonneau},
  {Ciardi}, {Collier Cameron}, {Cosentino}, {Crossfield}, {Damasso},
  {Dressing}, {Dumusque}, {Everett}, {Figueira}, {Fiorenzano}, {Gonzales},
  {Haywood}, {Harutyunyan}, {Hirsch}, {Howell}, {Johnson}, {Latham}, {Lopez},
  {Mayor}, {Micela}, {Molinari}, {Nascimbeni}, {Pepe}, {Phillips}, {Piotto},
  {Rice}, {Sasselov}, {S{\'e}gransan}, {Sozzetti}, {Udry}, \&
  {Watson}}]{Malavolta2018}
{Malavolta}, L., {Mayo}, A.~W., {Louden}, T., {et~al.} 2018, \aj, 155, 107

\bibitem[{{Mamajek} \& {Hillenbrand}(2008)}]{Mamajek}
{Mamajek}, E.~E., \& {Hillenbrand}, L.~A. 2008, \apj, 687, 1264

\bibitem[{MAST(2021)}]{mastdoi}
MAST. 2021, TESS Light Curves - All Sectors,  STScI/MAST,
  doi:10.17909/T9-NMC8-F686.
\newblock
  \url{http://archive.stsci.edu/doi/resolve/resolve.html?doi=10.17909/t9-nmc8-f686}

\bibitem[{{Masuda} {et~al.}(2022){Masuda}, {Petigura}, \& {Hall}}]{Masuda2022}
{Masuda}, K., {Petigura}, E.~A., \& {Hall}, O.~J. 2022, \mnras, 510, 5623

\bibitem[{{Mazeh} {et~al.}(2016){Mazeh}, {Holczer}, \& {Faigler}}]{Mazeh2016}
{Mazeh}, T., {Holczer}, T., \& {Faigler}, S. 2016, \aap, 589, A75

\bibitem[{{Metropolis} {et~al.}(1953){Metropolis}, {Rosenbluth}, {Rosenbluth},
  {Teller}, \& {Teller}}]{metropolis53}
{Metropolis}, N., {Rosenbluth}, A.~W., {Rosenbluth}, M.~N., {Teller}, A.~H., \&
  {Teller}, E. 1953, \jcp, 21, 1087

\bibitem[{{Millholland} \& {Spalding}(2020)}]{Millholland_USP}
{Millholland}, S.~C., \& {Spalding}, C. 2020, \apj, 905, 71

\bibitem[{{Murgas} {et~al.}(2022){Murgas}, {Nowak}, {Masseron}, {Parviainen},
  {Luque}, {Pall{\'e}}, {Korth}, {Carleo}, {Csizmadia}, {Esparza-Borges},
  {Alqasim}, {Cochran}, {Dai}, {Deeg}, {Gandolfi}, {Goffo}, {Kab{\'a}th},
  {Lam}, {Livingston}, {Muresan}, {Osborne}, {Persson}, {Serrano}, {Smith},
  {Van Eylen}, {Orell-Miquel}, {Hinkel}, {Gal{\'a}n}, {Puig-Subir{\`a}},
  {Stangret}, {Fukui}, {Kagetani}, {Narita}, {Ciardi}, {Boyle}, {Ziegler},
  {Brice{\~n}o}, {Law}, {Mann}, {Jenkins}, {Latham}, {Quinn}, {Ricker},
  {Seager}, {Shporer}, {Ting}, {Vanderspek}, \& {Winn}}]{hd20329}
{Murgas}, F., {Nowak}, G., {Masseron}, T., {et~al.} 2022, \aap, 668, A158

\bibitem[{{Murray-Clay} {et~al.}(2009){Murray-Clay}, {Chiang}, \&
  {Murray}}]{Murray-Clay}
{Murray-Clay}, R.~A., {Chiang}, E.~I., \& {Murray}, N. 2009, \apj, 693, 23

\bibitem[{{NASA Exoplanet Archive}(2019)}]{neadoi}
{NASA Exoplanet Archive}. 2019, Confirmed Planets Table,  IPAC,
  doi:10.26133/NEA1.
\newblock \url{https://catcopy.ipac.caltech.edu/dois/doi.php?id=10.26133/NEA1}

\bibitem[{{Neal}(2012)}]{neal12}
{Neal}, R.~M. 2012, arXiv e-prints, arXiv:1206.1901

\bibitem[{Newville {et~al.}(2014)Newville, Stensitzki, Allen, \&
  Ingargiola}]{LM}
Newville, M., Stensitzki, T., Allen, D.~B., \& Ingargiola, A. 2014, {LMFIT:
  Non-Linear Least-Square Minimization and Curve-Fitting for Python}, v0.8.0,
  Zenodo, doi:10.5281/zenodo.11813.
\newblock \url{https://doi.org/10.5281/zenodo.11813}

\bibitem[{{Owen} \& {Wu}(2017)}]{OwenWu2017}
{Owen}, J.~E., \& {Wu}, Y. 2017, \apj, 847, 29

\bibitem[{{Paegert} {et~al.}(2021){Paegert}, {Stassun}, {Collins}, {Pepper},
  {Torres}, {Jenkins}, {Twicken}, \& {Latham}}]{ticv82}
{Paegert}, M., {Stassun}, K.~G., {Collins}, K.~A., {et~al.} 2021, arXiv
  e-prints, arXiv:2108.04778

\bibitem[{pandas~development team(2020)}]{pandas}
pandas~development team, T. 2020, pandas-dev/pandas: Pandas, vlatest,  Zenodo,
  doi:10.5281/zenodo.3509134.
\newblock \url{https://doi.org/10.5281/zenodo.3509134}

\bibitem[{{Petigura}(2015)}]{Petigura2015}
{Petigura}, E.~A. 2015, PhD thesis, University of California, Berkeley

\bibitem[{{Rafikov}(2006)}]{Rafikov}
{Rafikov}, R.~R. 2006, \apj, 648, 666

\bibitem[{{Rajpaul} {et~al.}(2015){Rajpaul}, {Aigrain}, {Osborne}, {Reece}, \&
  {Roberts}}]{Rajpaul2015}
{Rajpaul}, V., {Aigrain}, S., {Osborne}, M.~A., {Reece}, S., \& {Roberts}, S.
  2015, \mnras, 452, 2269

\bibitem[{{Ribas} {et~al.}(2005){Ribas}, {Guinan}, {G{\"u}del}, \&
  {Audard}}]{Ribas}
{Ribas}, I., {Guinan}, E.~F., {G{\"u}del}, M., \& {Audard}, M. 2005, \apj, 622,
  680

\bibitem[{{Ricker} {et~al.}(2015){Ricker}, {Winn}, {Vanderspek}, {Latham},
  {Bakos}, {Bean}, {Berta-Thompson}, {Brown}, {Buchhave}, {Butler}, {Butler},
  {Chaplin}, {Charbonneau}, {Christensen-Dalsgaard}, {Clampin}, {Deming},
  {Doty}, {De Lee}, {Dressing}, {Dunham}, {Endl}, {Fressin}, {Ge}, {Henning},
  {Holman}, {Howard}, {Ida}, {Jenkins}, {Jernigan}, {Johnson}, {Kaltenegger},
  {Kawai}, {Kjeldsen}, {Laughlin}, {Levine}, {Lin}, {Lissauer}, {MacQueen},
  {Marcy}, {McCullough}, {Morton}, {Narita}, {Paegert}, {Palle}, {Pepe},
  {Pepper}, {Quirrenbach}, {Rinehart}, {Sasselov}, {Sato}, {Seager},
  {Sozzetti}, {Stassun}, {Sullivan}, {Szentgyorgyi}, {Torres}, {Udry}, \&
  {Villasenor}}]{TESS}
{Ricker}, G.~R., {Winn}, J.~N., {Vanderspek}, R., {et~al.} 2015, Journal of
  Astronomical Telescopes, Instruments, and Systems, 1, 014003

\bibitem[{{Rouan} {et~al.}(2011){Rouan}, {Deeg}, {Demangeon}, {Samuel},
  {Cavarroc}, {Fegley}, \& {L{\'e}ger}}]{Rouan2011}
{Rouan}, D., {Deeg}, H.~J., {Demangeon}, O., {et~al.} 2011, \apjl, 741, L30

\bibitem[{Saar \& Donahue(1997)}]{Saar1997}
Saar, S.~H., \& Donahue, R.~A. 1997, The Astrophysical Journal, 485, 319.
\newblock \url{https://dx.doi.org/10.1086/304392}

\bibitem[{Salvatier {et~al.}(2016)Salvatier, Wiecki, \& Fonnesbeck}]{pymc3}
Salvatier, J., Wiecki, T.~V., \& Fonnesbeck, C. 2016, PeerJ Computer Science,
  2, doi:10.7717/peerj-cs.55

\bibitem[{Sanchis-Ojeda {et~al.}(2014)Sanchis-Ojeda, Rappaport, Winn, Kotson,
  Levine, \& Mellah}]{SanchisOjeda2014}
Sanchis-Ojeda, R., Rappaport, S., Winn, J.~N., {et~al.} 2014, The Astrophysical
  Journal, 787, 47.
\newblock \url{https://dx.doi.org/10.1088/0004-637X/787/1/47}

\bibitem[{{Sanchis-Ojeda} {et~al.}(2013){Sanchis-Ojeda}, {Rappaport}, {Winn},
  {Levine}, {Kotson}, {Latham}, \& {Buchhave}}]{Sanchis2013}
{Sanchis-Ojeda}, R., {Rappaport}, S., {Winn}, J.~N., {et~al.} 2013, \apj, 774,
  54

\bibitem[{{Scott} {et~al.}(2021){Scott}, {Howell}, {Gnilka}, {Stephens},
  {Salinas}, {Matson}, {Furlan}, {Horch}, {Everett}, {Ciardi}, {Mills}, \&
  {Quigley}}]{Alopeke}
{Scott}, N.~J., {Howell}, S.~B., {Gnilka}, C.~L., {et~al.} 2021, Frontiers in
  Astronomy and Space Sciences, 8, 138

\bibitem[{{Seager} \& {Mall{\'e}n-Ornelas}(2003)}]{Seager}
{Seager}, S., \& {Mall{\'e}n-Ornelas}, G. 2003, \apj, 585, 1038

\bibitem[{{STScI}(2018)}]{ticdoi}
{STScI}. 2018, TESS Input Catalog and Candidate Target List,  STScI/MAST,
  doi:10.17909/FWDT-2X66.
\newblock
  \url{http://archive.stsci.edu/doi/resolve/resolve.html?doi=10.17909/fwdt-2x66}

\bibitem[{{Tayar} {et~al.}(2022){Tayar}, {Claytor}, {Huber}, \& {van
  Saders}}]{Tayar2022}
{Tayar}, J., {Claytor}, Z.~R., {Huber}, D., \& {van Saders}, J. 2022, \apj,
  927, 31

\bibitem[{{Tsantaki} {et~al.}(2020){Tsantaki}, {Andreasen}, \&
  {Teixeira}}]{Tsantaki2020}
{Tsantaki}, M., {Andreasen}, D., \& {Teixeira}, G. 2020, The Journal of Open
  Source Software, 5, 2048

\bibitem[{{Tsantaki} {et~al.}(2018){Tsantaki}, {Andreasen}, {Teixeira},
  {Sousa}, {Santos}, {Delgado-Mena}, \& {Bruzual}}]{Tsantaki2018}
{Tsantaki}, M., {Andreasen}, D.~T., {Teixeira}, G.~D.~C., {et~al.} 2018,
  \mnras, 473, 5066

\bibitem[{{Tu} {et~al.}(2015){Tu}, {Johnstone}, {G{\"u}del}, \&
  {Lammer}}]{Tu2015}
{Tu}, L., {Johnstone}, C.~P., {G{\"u}del}, M., \& {Lammer}, H. 2015, \aap, 577,
  L3

\bibitem[{{Vehtari} {et~al.}(2021){Vehtari}, {Gelman}, {Simpson}, {Carpenter},
  \& {B{\"u}rkner}}]{vehtari21}
{Vehtari}, A., {Gelman}, A., {Simpson}, D., {Carpenter}, B., \& {B{\"u}rkner},
  P.-C. 2021, Bayesian Analysis, 16, 667

\bibitem[{Virtanen {et~al.}(2020)Virtanen, Gommers, Oliphant, Haberland, Reddy,
  Cournapeau, Burovski, Peterson, Weckesser, Bright, {van der Walt}, Brett,
  Wilson, Millman, Mayorov, Nelson, Jones, Kern, Larson, Carey, Polat, Feng,
  Moore, {VanderPlas}, Laxalde, Perktold, Cimrman, Henriksen, Quintero, Harris,
  Archibald, Ribeiro, Pedregosa, {van Mulbregt}, \& {SciPy 1.0
  Contributors}}]{scipy}
Virtanen, P., Gommers, R., Oliphant, T.~E., {et~al.} 2020, NatMe, 17, 261

\bibitem[{{Vogt} {et~al.}(1994){Vogt}, {Allen}, {Bigelow}, {Bresee}, {Brown},
  {Cantrall}, {Conrad}, {Couture}, {Delaney}, {Epps}, {Hilyard}, {Hilyard},
  {Horn}, {Jern}, {Kanto}, {Keane}, {Kibrick}, {Lewis}, {Osborne},
  {Pardeilhan}, {Pfister}, {Ricketts}, {Robinson}, {Stover}, {Tucker}, {Ward},
  \& {Wei}}]{HIRES}
{Vogt}, S.~S., {Allen}, S.~L., {Bigelow}, B.~C., {et~al.} 1994, in Society of
  Photo-Optical Instrumentation Engineers (SPIE) Conference Series, Vol. 2198,
  Instrumentation in Astronomy VIII, ed. D.~L. {Crawford} \& E.~R. {Craine},
  362

\bibitem[{{Wang} \& {Dai}(2018)}]{WangDai2018}
{Wang}, L., \& {Dai}, F. 2018, \apj, 860, 175

\bibitem[{{Weiss} \& {Marcy}(2014)}]{Weiss2014}
{Weiss}, L.~M., \& {Marcy}, G.~W. 2014, \apjl, 783, L6

\bibitem[{{Weiss} {et~al.}(2021){Weiss}, {Dai}, {Huber}, {Brewer}, {Collins},
  {Ciardi}, {Matthews}, {Ziegler}, {Howell}, {Batalha}, {Crossfield},
  {Dressing}, {Fulton}, {Howard}, {Isaacson}, {Kane}, {Petigura}, {Robertson},
  {Roy}, {Rubenzahl}, {Twicken}, {Claytor}, {Stassun}, {MacDougall}, {Chontos},
  {Giacalone}, {Dalba}, {Mocnik}, {Hill}, {Beard}, {Akana Murphy}, {Rosenthal},
  {Behmard}, {Van Zandt}, {Lubin}, {Kosiarek}, {Lund}, {Christiansen},
  {Matson}, {Beichman}, {Schlieder}, {Gonzales}, {Brice{\~n}o}, {Law}, {Mann},
  {Collins}, {Evans}, {Fukui}, {Jensen}, {Murgas}, {Narita}, {Palle},
  {Parviainen}, {Schwarz}, {Tan}, {Acton}, {Bryant}, {Chaushev}, {Gill},
  {Eigm{\"u}ller}, {Jenkins}, {Ricker}, {Seager}, \& {Winn}}]{Weiss561}
{Weiss}, L.~M., {Dai}, F., {Huber}, D., {et~al.} 2021, \aj, 161, 56

\bibitem[{{Zeng} {et~al.}(2016){Zeng}, {Sasselov}, \& {Jacobsen}}]{Zeng2016}
{Zeng}, L., {Sasselov}, D.~D., \& {Jacobsen}, S.~B. 2016, \apj, 819, 127

\bibitem[{{Zieba} {et~al.}(2022){Zieba}, {Zilinskas}, {Kreidberg}, {Nguyen},
  {Miguel}, {Cowan}, {Pierrehumbert}, {Carone}, {Dang}, {Hammond}, {Louden},
  {Lupu}, {Malavolta}, \& {Stevenson}}]{Zieba2022}
{Zieba}, S., {Zilinskas}, M., {Kreidberg}, L., {et~al.} 2022, \aap, 664, A79

\end{thebibliography}
\bibliographystyle{aasjournal}

\appendix

\begin{figure*}[b!]
    \centering
    \includegraphics[width=\textwidth]{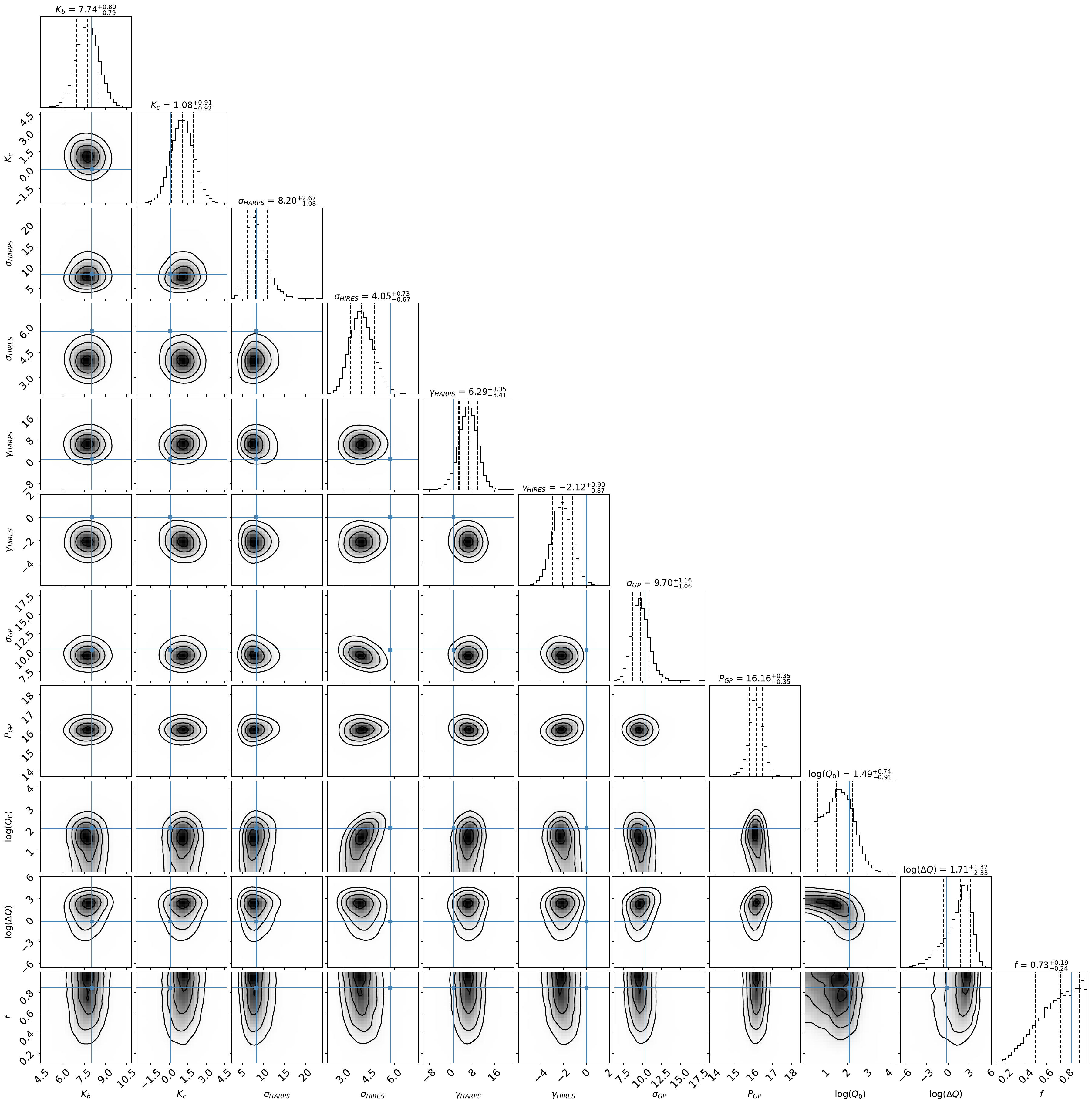}
    \caption{Full posterior distributions for the RV model described in Section~\ref{sec:rvmodel}. The blue lines denote the MAP values.}
    \label{fig:cornerplot}
\end{figure*}

Here we present the posterior distributions for the full RV model (Figure~\ref{fig:cornerplot}), as well as the table of RV observations analyzed in Section~\ref{sec:rvs} (Table~\ref{tab:rvdata}).

\startlongtable
\begin{deluxetable*}{lllllr}
\tablecaption{TOI-1347 RVs and S-Indices \label{tab:rvdata}}
\tablehead{\colhead{Time} & \colhead{RV\tablenotemark{a}} & \colhead{RV Error} & \colhead{$S_\mathrm{HK}$ Index} & \colhead{$S_\mathrm{HK}$ Index Error} & \colhead{Instrument} \\ \colhead{(BJD-2457000)} & \colhead{(m/s)} & \colhead{(m/s)} & \colhead{} & \colhead{} & \colhead{}}
\startdata
2458974.658616  &  -2.385  & 2.91  & --    & --     & HARPS-N \\
2458976.56778   &  3.675   & 2.77  & --    & --     & HARPS-N \\
2458977.647243  &  1.645   & 2.3   & --    & --     & HARPS-N \\
2458978.664451  &  1.735   & 1.84  & --    & --     & HARPS-N \\
2458979.661071  &  15.635  & 2.56  & --    & --     & HARPS-N \\
2458983.624618  &  2.145   & 2.92  & --    & --     & HARPS-N \\
2458985.688595  &  15.945  & 2.17  & --    & --     & HARPS-N \\
2458987.721405  &  8.195   & 2.07  & --    & --     & HARPS-N \\
2458988.675995  &  -1.315  & 1.57  & --    & --     & HARPS-N \\
2458989.689122  &  12.705  & 2.12  & --    & --     & HARPS-N \\
2458990.665386  &  -6.745  & 1.76  & --    & --     & HARPS-N \\
2458991.669197  &  1.215   & 3.26  & --    & --     & HARPS-N \\
2458992.622382  &  -1.215  & 4.22  & --    & --     & HARPS-N \\
2458993.586943  &  -8.595  & 2.12  & --    & --     & HARPS-N \\
2459011.871925  &  27.023  & 1.73  & 0.291 & 0.001  & HIRES \\
2459011.945652  &  31.576  & 1.75  & 0.285 & 0.001  & HIRES \\
2459011.999298  &  35.149  & 1.85  & 0.287 & 0.001  & HIRES \\
2459028.000963  &  21.648  & 1.49  & 0.285 & 0.001  & HIRES \\
2459034.871693  &  1.685   & 1.74  & 0.284 & 0.001  & HIRES \\
2459036.801673  &  5.49    & 1.85  & 0.281 & 0.001  & HIRES \\
2459040.959633  &  7.374   & 1.88  & 0.277 & 0.001  & HIRES \\
2459043.067716  &  -14.481 & 1.69  & 0.278 & 0.001  & HIRES \\
2459073.019955  &  -4.758  & 1.77  & 0.263 & 0.001  & HIRES \\
2459079.02133   &  13.265  & 1.94  & 0.271 & 0.001  & HIRES \\
2459086.759188  &  8.866   & 2.45  & 0.261 & 0.001  & HIRES \\
2459094.963738  &  -0.889  & 1.81  & 0.256 & 0.001  & HIRES \\
2459099.906082  &  -3.443  & 2.08  & 0.258 & 0.001  & HIRES \\
2459101.74781   &  -6.744  & 1.72  & 0.275 & 0.001  & HIRES \\
2459102.001018  &  0.082   & 1.95  & 0.274 & 0.001  & HIRES \\
2459114.730454  &  17.792  & 1.68  & 0.277 & 0.001  & HIRES \\
2459114.919621  &  11.108  & 1.97  & 0.277 & 0.001  & HIRES \\
2459115.767787  &  4.19    & 1.76  & 0.281 & 0.001  & HIRES \\
2459115.851596  &  -2.33   & 1.81  & 0.278 & 0.001  & HIRES \\
2459117.73074   &  -17.608 & 1.7   & 0.276 & 0.001  & HIRES \\
2459117.892083  &  -8.786  & 1.65  & 0.274 & 0.001  & HIRES \\
2459118.725156  &  -2.82   & 1.71  & 0.275 & 0.001  & HIRES \\
2459119.822351  &  3.015   & 1.62  & 0.262 & 0.001  & HIRES \\
2459120.849961  &  -0.745  & 1.84  & 0.261 & 0.001  & HIRES \\
2459121.893938  &  -10.875 & 1.62  & 0.263 & 0.001  & HIRES \\
2459122.813237  &  -8.108  & 1.67  & 0.264 & 0.001  & HIRES \\
2459123.723648  &  -7.533  & 1.58  & 0.264 & 0.001  & HIRES \\
2459189.746637  &  -25.975 & 2.66  & 0.279 & 0.001  & HIRES \\
2459297.143477  &  -1.254  & 2.29  & 0.319 & 0.001  & HIRES \\
2459373.951805  &  9.778   & 1.68  & 0.306 & 0.001  & HIRES \\
2459377.047988  &  -12.886 & 1.79  & 0.296 & 0.001  & HIRES \\
2459377.869358  &  -18.981 & 1.72  & 0.296 & 0.001  & HIRES \\
2459379.121054  &  7.817   & 1.65  & 0.29  & 0.001  & HIRES \\
2459379.945803  &  -1.632  & 1.65  & 0.282 & 0.001  & HIRES \\
2459382.038048  &  -28.729 & 2.01  & 0.268 & 0.001  & HIRES \\
2459382.957833  &  -29.687 & 1.89  & 0.278 & 0.001  & HIRES \\
2459385.051386  &  0.288   & 1.78  & 0.285 & 0.001  & HIRES \\
2459385.877791  &  7.743   & 1.73  & 0.289 & 0.001  & HIRES \\
2459388.847531  &  8.234   & 1.79  & 0.309 & 0.001  & HIRES \\
2459389.113887  &  20.363  & 1.85  & 0.313 & 0.001  & HIRES \\
2459389.840061  &  12.641  & 1.64  & 0.311 & 0.001  & HIRES \\
2459390.10277   &  7.56    & 1.79  & 0.31  & 0.001  & HIRES \\
2459396.049221  &  -11.462 & 1.87  & 0.281 & 0.001  & HIRES \\
2459405.116336  &  8.808   & 1.68  & 0.309 & 0.001  & HIRES \\
2459405.952026  &  9.326   & 1.77  & 0.315 & 0.001  & HIRES \\
2459406.101749  &  33.565  & 1.79  & 0.31  & 0.001  & HIRES \\
2459407.905698  &  27.179  & 1.8   & 0.0   & 0.001  & HIRES \\
2459413.001295  &  8.237   & 1.94  & 0.284 & 0.001  & HIRES \\
2459413.964231  &  -1.111  & 1.9   & 0.278 & 0.001  & HIRES \\
2459414.118005  &  -18.948 & 2.63  & 0.239 & 0.001  & HIRES \\
2459420.770145  &  8.008   & 1.68  & 0.292 & 0.001  & HIRES \\
2459441.889385  &  -8.624  & 1.95  & 0.311 & 0.001  & HIRES \\
2459442.029629  &  -18.863 & 1.85  & 0.305 & 0.001  & HIRES \\
2459443.913165  &  -20.67  & 1.83  & 0.302 & 0.001  & HIRES \\
2459446.031598  &  0.282   & 1.75  & 0.283 & 0.001  & HIRES \\
2459448.761484  &  -3.308  & 2.23  & 0.281 & 0.001  & HIRES \\
2459449.003801  &  -10.891 & 2.09  & 0.277 & 0.001  & HIRES \\
2459449.860035  &  -10.669 & 1.84  & 0.278 & 0.001  & HIRES \\
2459451.981593  &  3.237   & 1.83  & 0.273 & 0.001  & HIRES \\
2459452.83527   &  14.859  & 1.76  & 0.283 & 0.001  & HIRES \\
2459455.789937  &  14.152  & 1.61  & 0.303 & 0.001  & HIRES \\
2459456.77851   &  4.719   & 1.99  & 0.311 & 0.001  & HIRES \\
2459469.778004  &  7.652   & 1.7   & 0.291 & 0.001  & HIRES \\
2459470.793047  &  10.455  & 1.6   & 0.299 & 0.001  & HIRES \\
2459472.733849  &  -5.021  & 1.58  & 0.301 & 0.001  & HIRES \\
2459474.860635  &  -0.301  & 1.87  & 0.301 & 0.001  & HIRES \\
2459475.765816  &  6.237   & 1.91  & 0.296 & 0.001  & HIRES \\
2459481.868763  &  -21.987 & 1.88  & 0.286 & 0.001  & HIRES \\
2459482.872647  &  -13.532 & 1.72  & 0.285 & 0.001  & HIRES \\
2459483.768718  &  -10.044 & 1.85  & 0.284 & 0.001  & HIRES \\
2459484.888563  &  17.3    & 1.8   & 0.294 & 0.001  & HIRES \\
2459497.736124  &  -12.074 & 1.92  & 0.283 & 0.001  & HIRES \\
2459498.730087  &  -4.543  & 1.77  & 0.284 & 0.001  & HIRES \\
2459498.878826  &  -10.565 & 1.89  & 0.286 & 0.001  & HIRES \\
2459502.866887  &  0.693   & 1.82  & 0.316 & 0.001  & HIRES \\
2459503.799484  &  13.319  & 1.74  & 0.307 & 0.001  & HIRES \\
2459504.809106  &  12.162  & 2.11  & 0.316 & 0.001  & HIRES \\
2459506.714657  &  -6.684  & 1.96  & 0.322 & 0.001  & HIRES \\
2459508.769583  &  1.738   & 1.86  & 0.31  & 0.001  & HIRES \\
2459509.853591  &  -7.617  & 1.97  & 0.302 & 0.001  & HIRES \\
2459513.787074  &  -12.625 & 2.06  & 0.29  & 0.001  & HIRES \\
2459516.749338  &  -5.973  & 1.84  & 0.288 & 0.001  & HIRES \\
2459541.764685  &  8.434   & 2.21  & 0.306 & 0.001  & HIRES \\
2459542.750956  &  -4.746  & 2.14  & 0.31  & 0.001  & HIRES \\
2459543.709322  &  -4.926  & 2.34  & 0.299 & 0.001  & HIRES \\
2459545.717882  &  -20.888 & 2.51  & 0.292 & 0.001  & HIRES \\
2459546.758504  &  -10.908 & 2.34  & 0.292 & 0.001  & HIRES \\
2459681.066484  &  -11.429 & 2.97  & 0.308 & 0.001  & HIRES \\
2459690.076369  &  26.327  & 2.2   & 0.326 & 0.001  & HIRES \\
2459691.125173  &  21.194  & 2.28  & 0.33  & 0.001  & HIRES \\
2459694.098406  &  -6.466  & 3.98  & 0.338 & 0.001  & HIRES \\
2459695.041332  &  -8.45   & 2.31  & 0.334 & 0.001  & HIRES \\
2459708.022865  &  20.552  & 2.18  & 0.336 & 0.001  & HIRES \\
2459710.08107   &  3.64    & 2.11  & 0.335 & 0.001  & HIRES \\
2459712.956673  &  -5.968  & 2.1   & 0.308 & 0.001  & HIRES \\
2459716.048527  &  -9.38   & 2.01  & 0.297 & 0.001  & HIRES \\
2459738.012297  &  -1.455  & 1.86  & 0.316 & 0.001  & HIRES \\
2459738.903643  &  9.585   & 1.9   & 0.326 & 0.001  & HIRES \\
2459739.110148  &  20.248  & 1.84  & 0.337 & 0.001  & HIRES \\
2459739.881402  &  14.875  & 1.83  & 0.335 & 0.001  & HIRES \\
2459740.912726  &  22.59   & 1.81  & 0.338 & 0.001  & HIRES \\
2459741.96611   &  4.237   & 1.78  & 0.328 & 0.001  & HIRES \\
2459742.920385  &  -8.773  & 2.08  & 0.33  & 0.001  & HIRES \\
2459745.878305  &  -10.306 & 1.83  & 0.295 & 0.001  & HIRES \\
2459745.911268  &  -11.487 & 1.75  & 0.301 & 0.001  & HIRES \\
2459747.905515  &  -6.925  & 1.71  & 0.288 & 0.001  & HIRES \\
2459748.075272  &  -15.008 & 1.75  & 0.287 & 0.001  & HIRES \\
2459748.985451  &  -24.499 & 1.81  & 0.286 & 0.001  & HIRES \\
2459749.881961  &  -19.976 & 1.77  & 0.283 & 0.001  & HIRES \\
2459756.993647  &  24.379  & 1.78  & 0.318 & 0.001  & HIRES \\
2459760.025542  &  -12.004 & 1.92  & 0.313 & 0.001  & HIRES \\
2459766.003617  &  -26.032 & 1.73  & 0.281 & 0.001  & HIRES \\
2459769.01945   &  9.678   & 1.8   & 0.286 & 0.001  & HIRES \\
2459770.019978  &  4.399   & 1.58  & 0.298 & 0.001  & HIRES \\
2459771.904391  &  4.986   & 1.85  & 0.314 & 0.001  & HIRES \\
2459776.819221  &  -23.025 & 2.38  & 0.304 & 0.001  & HIRES \\
2459780.003495  &  -0.199  & 1.76  & 0.293 & 0.001  & HIRES \\
2459780.996733  &  -8.457  & 1.84  & 0.291 & 0.001  & HIRES \\
2459785.993673  &  12.388  & 1.79  & 0.292 & 0.001  & HIRES \\
2459787.063914  &  4.716   & 1.84  & 0.299 & 0.001  & HIRES \\
\enddata
\tablenotetext{a}{A median value of $-13965.395$~{\kms} has been subtracted from the HARPS-N RVs.}
\end{deluxetable*}

\end{document}